\documentclass[preprint,12pt,authoryear]{elsarticle}
\usepackage[margin=2cm]{geometry}
\usepackage[utf8]{inputenc}
\usepackage[english]{babel}
\usepackage{natbib}
\usepackage{graphicx}
\graphicspath{ {figures/} }
\usepackage{siunitx}
\usepackage{amsmath}
\usepackage{amssymb}
\usepackage{subcaption}
\usepackage{multirow}
\usepackage{booktabs}
\usepackage{tabularx}
\usepackage[colorlinks=true,linkcolor=black, citecolor=blue, urlcolor=blue]{hyperref}
\usepackage[nameinlink, capitalize]{cleveref}
\usepackage{xurl}

\usepackage{mathtools}

\usepackage{makecell}

\begin{document}

\begin{frontmatter}
\title{Bounded-METANET: A new discrete-time second-order macroscopic traffic flow model for bounded speed}

\author[inst1]{Weiming Zhao}
\ead{weiming.zhao@uq.edu.au, joyfig07@gmail.com}

\author[inst2,inst3]{Claudio Roncoli}
\ead{claudio.roncoli@kuleuven.be}

\author[inst1]{Mehmet Yildirimoglu}
\ead{m.yildirimoglu@uq.edu.au}

\affiliation[inst1]{
    organization={School of Civil Engineering},
    university={The University of Queensland,},
    city={Brisbane QLD 4072},
    country={Australia}
}

\affiliation[inst2]{
    organization={Centre for Industrial Management / Traffic and Infrastructure},
    university={KU Leuven,},
    city={Leuven 3001},
    country={Belgium}
}

\affiliation[inst3]{
    organization={Department of Built Environment},
    university={Aalto University,},
    city={Espoo 02150},
    country={Finland}
}

\begin{abstract}

Macroscopic traffic flow models are essential for analysing traffic dynamics in highways and urban roads. While second-order models like METANET capture non-equilibrium traffic states, they often produce unrealistic speed predictions, such as negative values or speeds above the free-flow limit, which limits their reliability in traffic management. To overcome these limitations, we introduce Bounded-METANET, a new discrete-time second-order model that refines METANET’s speed update equation by removing the convection term and adding a virtual density mechanism to reflect anticipation and merging effects. This ensures that speeds stay bounded between zero and the free-flow speed, simplifying calibration and boosting usability. Validated with SUMO simulations and real-world German highway data, Bounded-METANET accurately captures non-equilibrium flow and the capacity drop phenomenon, and outperforms METANET in estimating the fundamental diagram under congestion. It achieves lower RMSE for speed and density in noise-free simulation data and better flow estimation in real-world data, though METANET edges out in speed RMSE. Unlike METANET, which can produce erratic shockwave speeds and flow errors, Bounded-METANET delivers consistent, realistic predictions. This makes it a promising tool for traffic modelling and control in various scenarios.

\end{abstract}

\begin{keyword}
macroscopic traffic flow modelling \sep METANET \sep freeway traffic flow \sep second-order traffic flow model \sep fundamental diagram
\end{keyword}

\end{frontmatter}

\section{Introduction}

Macroscopic traffic flow modelling is a widely used tool in the simulation and optimisation of traffic networks, including highway networks and urban road networks. Macroscopic models are capable of simulating the dynamics of aggregated traffic flow variables, namely link speed, link density, and link flow. In the context of traffic simulation, they are more scalable than microscopic traffic flow models and can be used to efficiently predict traffic states in large networks over extended periods of time. In traffic control, macroscopic models are typically used to determine dynamic speed limits and regulate ramp metering on motorways \citep{papageorgiouFreeway2002, carlsonOptimal2010}.

Macroscopic models can be divided into first and second order models. \citet{lighthill1955kinematic} and \citet{richards1956Shock} proposed the first continuum model to describe traffic flow, also known as the LWR model. The conservation law is defined by the following equation:
\begin{equation}
    \frac{\partial \rho(x,t)}{\partial t} + \frac{\partial q(x, t)}{\partial x} = 0
\end{equation}
\citet{daganzoCell1994} proposed the Cell Transmission Model (CTM), which is the Godunov discrete version of the LWR model. Early attempts to apply the Godunov discretisation of the LWR model are reported in \citet{lebacqueSemimacroscopic1984, lebacqueGodunov1996}. 
First-order models assume equilibrium traffic states, where speed aligns with density via the fundamental diagram. However, they critically rely on the physically unrealistic assumption of instantaneous and unbounded acceleration, a contradiction to the dynamics of real-world vehicles. To address this shortcoming, several bounded acceleration extensions to the LWR model have been developed. Notable contributions include the seminal bounded acceleration model by \citet{lebacqueFinite1997}, Lebacque's subsequent two-phase approach \citep{lebacqueTwoPhase2003}, and further theoretical advancements by \citep{leclercqBounded2007} and \citep{jinBounded2018}.

Differently, in second-order models, to simulate the non-equilibrium traffic state, another conservation equation is introduced that models the local speed by an acceleration equation, as follows:
\begin{equation}
    \frac{d V(x,t)}{dt} = \frac{\partial V(x,t)}{\partial t} + V(x,t) \frac{\partial V(x, t)}{\partial x} = f(\rho(x,t), V(x,t), \frac{\partial \rho(x,t)}{\partial x}, \frac{\partial V(x,t)}{\partial x})
\end{equation}
leading to the general second order model
\begin{equation}
\begin{cases}
\dfrac{\partial \rho(x,t)}{\partial t} + \dfrac{\partial q(x, t)}{\partial x} = 0 \\ 
\dfrac{\partial V(x,t)}{\partial t} + V(x,t) \dfrac{\partial V(x, t)}{\partial x} = f(\rho(x,t), V(x,t), \dfrac{\partial \rho(x,t)}{\partial x}, \dfrac{\partial V(x,t)}{\partial x})
\end{cases}
\end{equation}
The Payne model \citep{payne1971model} is the prototype of the second order models. The acceleration function is the sum of relaxation and anticipation terms. The problem with this model is that the characteristic speed is sometimes faster than the macroscopic vehicle speed, which means that the information travels faster than the vehicle, and the future traffic state is determined by the upstream state \citep{daganzoRequiem1995}. A discrete version of the second-order model derived from Payne's model is proposed by \citet{papageorgiouMacroscopic1989} and is called METANET, which has been further developed to deal with different network layouts such as off-ramps, on-ramps and lane drops. It has been used to model network traffic flow \citep{kotsialosTraffic2002}, motorway control and management such as ramp metering \citep{papamichailCoordinated2010}, variable speed limits \citep{carlsonOptimal2010}, and motorway-to-motorway control \citep{carlsonOptimal2010b}.

METANET has been extensively validated using real traffic data such as in Amsterdam (Netherlands) \citep{kotsialosTraffic2002}, Athens (Greece) \citep{spiliopoulouMacroscopic2014}, Delft (Netherlands) \citep{ngoduyAutomated2003}, Frankfurt (Germany) \citep{mohammadianPerformance2021}, Shanghai (China) \citep{wangMacroscopic2022}, Sheffield (UK) \citep{pooleSecond2016}. Complex traffic environments are considered in the verification, including congestion due to off-ramps \citep{spiliopoulouMacroscopic2014}, congestion due to accidents \citep{mohammadianPerformance2021}, adverse weather conditions \citep{wangMacroscopic2022}.
Comparative studies of traffic flow models, such as those by \citet{papageorgiouMacroscopic1989} (e.g., LWR, Payne, and METANET), \citet{kontorinakiFirstorder2017} (e.g., METANET and CTM), which aimed to assess approaches to creating capacity drop in first-order models, and \citet{mohammadianPerformance2021} (e.g., CTM, LWR, FASTLANE, METANET, and three other second-order models), consistently identify METANET as delivering the highest calibration performance.

However, METANET has a significant problem: it does not ensure that speeds remain within a reasonable range between zero and the maximum allowed speed. In fact, during METANET simulations, the link speed can become negative or exceed the maximum speed. This problem is not unique to METANET, and many other second-order traffic flow models have a similar limitation, such as Aw--Rascle--Zhang (ARZ) ~\citep{awResurrection2000, zhangnonequilibrium2002}. Developing a numerical method for the ARZ model that satisfies bounds on traffic variables such as speed and density, also known as bound-preserving (BP) numerical methods, is very challenging. The Godunov scheme can produce spurious oscillations near contact discontinuities in the Aw-Rascle model and violate the reasonable boundaries of speed. \citet{chalonsTransportequilibrium2007} proposed a method by combining the Godunov method with a random sampling strategy that satisfies the maximum principle of speed, i.e. \(v \le  v_{\textrm{max}}\). However, due to the random sampling, the conservation of density does not hold. \citet{betancourtrandom2018} verified that there is no strictly conservative BP method for the ARZ model that can satisfy the maximum principle of speed. \citet{chenBoundpreserving2025} developed a numerical BP method that preserves the minimum principle of speed, and the positivity of density. While it does not explicitly enforce the speed maximum principle, it maintains an alternative upper bound through the sum of speed and pressure.

The literature contains few extensions to METANET, with each addressing specific aspects of traffic modelling. \citet{pasqualeTwoclass2015} developed a multi-class METANET model where each vehicle class exhibits class-specific dynamics for density, speed, and on-ramp queues, while capturing vehicle interactions through a composite total density. Similarly, \citet{kotsialosvarying2021} proposed a multi-class METANET variant where fundamental diagram parameters, speed equations, and queue dynamics adapt based on traffic composition. More recently, \citet{huangIMETANET2024} introduced I-METANET, specifically designed to model incident traffic flow by modifying METANET's speed update equations to account for: incident-induced merging behaviour, interactions between upstream flows and downstream incident-created queue waves, and the combined effects of ramp flows and incident queues in interchange segments.

Nevertheless, the issue of unreasonable speed values in METANET has received little attention in the literature. A practical solution often used is to introduce a saturation function into the speed output, which ensures that the speed remains within a pre-specified range; however, this does not guarantee that the simulation in subsequent steps is an accurate representation of the expected result, as the speed may exhibit abrupt changes. It is challenging to ensure that the speed after the saturation function is an accurate representation of the real speed, even when all parameters are calibrated. This is because only a very limited number of instances where the saturation function is active are evaluated during the calibration process. If an abnormal speed that never occurs in the calibration process is present in the simulation, the resulting output is inherently uncertain and cannot be trusted.

This paper proposes a novel well-behaved and inherently bounded second-order macroscopic traffic flow model. The proposed model, inspired by METANET, addresses its limitations by ensuring link speeds remain within physically feasible ranges, never falling below zero or exceeding the free-flow speed. This feature facilitates calibration and enhances the model’s reliability by eliminating the need to handle unrealistic traffic variables during both calibration and application phases. Our reformulation removes the convection term from METANET’s speed equation, which decouples future speed dynamics from upstream traffic state dependencies. Key innovations of this approach include: 1)~redefined anticipation and merging effects: anticipation, merging and lane-drop impacts are mathematically embedded into the relaxation term via a virtual density mechanism, which ensures bounded speeds while preserving dynamic traffic behaviour; and 2)~intrinsic speed bounding: the model's formulation inherently restricts speeds to a pre-defined range~\([0, v_{\textrm{free}}]\), eliminating non-physical states without requiring ad-hoc corrections during simulation or control applications.

The remainder of the paper is organized as follows. \cref{sec:methodology} introduces the METANET model and proposes the Bounded-METANET model. It further presents a theoretical analysis of link speed and density boundaries to highlight the properties and advantages of the proposed method. \cref{sec:numerical_results} evaluates the performance of METANET, Bounded-METANET, and the first-order CTM through two numerical experiments: (1)~using simulated benchmark data and (2)~employing real-world traffic data collected via loop detectors on a motorway. Finally, \cref{sec:conclusions} summarises the key findings and concludes the paper.

\section{Methodology}~\label{sec:methodology}

This section first introduces the fundamentals of the METANET model and subsequently presents the proposed Bounded-METANET model. A theoretical analysis of the speed and density boundaries is conducted to demonstrate the key properties of the Bounded-METANET model.

\subsection{METANET}

\begin{figure}[tb]
    \centering
    \medskip
    \centering\includegraphics[width=.6\linewidth]{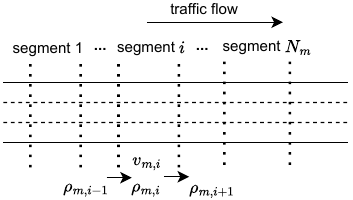}
    \caption{The link is divided into several segments.} \label{fig:metanet_variable}
\end{figure}

METANET is a macroscopic, discrete-time, second-order traffic flow model. A road network is divided into several links with a homogeneous layout and traffic composition. Each link is divided into several segments of approximately 300-500 m. A typical network layout is shown in \cref{fig:metanet_variable}. The simulation time step is typically 5-15 s. To ensure numerical stability, the Courant--Friedrichs--Lewy (CFL) condition must be satisfied: vehicles should not travel more than one segment in a simulation step \citep{courantUeber1928}. 

The METANET equations are as follows.
\begin{equation} \label{eq:density_update}
    \rho_{m,i}(k+1) = \rho_{m,i}(k) + \dfrac{T}{L_m \lambda_m} \left( q_{m,i-1}(k) - q_{m,i}(k) + q_o(k) - q_s(k) \right)
\end{equation}
\begin{equation} \label{eq:speed_update}
\begin{aligned}
    v_{m,i}(k+1) = & v_{m,i}(k) + \underbrace{\frac{T}{\tau} \left( V[\rho_{m,i}(k)] - v_{m,i}(k) \right)}_{\textrm{relaxation term}} + \\
   & \underbrace{\frac{T}{L_m} v_{m,i}(k) \left( v_{m,i-1}(k) - v_{m,i}(k) \right)}_{\textrm{convection term}} - \underbrace{\frac{\eta T}{\tau L_m} \frac{\rho_{m,i+1}(k) - \rho_{m,i}(k)}{\rho_{m,i}(k)+ \kappa}}_{\textrm{anticipation term}}
\end{aligned}
\end{equation}
\begin{equation}~\label{eq:flow}
q_{m,i}(k) = \rho_{m,i}(k) v_{m,i}(k)
\end{equation}
\begin{equation} \label{eq:fundamental_diagram}
V[\rho_{m,i}(k)] = v_{\mathrm{free},m} \exp{ \left\{ -\frac{1}{a_m} \left( \frac{\rho_{m,i}(k)}{\rho_{\mathrm{crit},m}}\right)^{a_m} \right\}}
\end{equation}
\begin{equation} \label{eq:queue_length_udpate}
    w_o(k+1) = w_o(k) + T(d_o(k) - q_o(k))
\end{equation}
\begin{equation} \label{eq:node_outflow}
    q_o(k) = \min \left[ d_o(k) + \frac{w_o(k)}{T}, C_o, C_o\left( \frac{\rho_{\max} - \rho_{m,i}(k)}{\rho_{\max} - \rho_{\mathrm{crit},m}} \right)  \right]
\end{equation}
where $T$ denotes the duration of the simulation time step, $q_{m,i}(k), \rho_{m,i}(k)$, and $v_{m,i}(k)$ denote the flow, density, and speed of link \(m\) segment \(i\) at time step \(k\), respectively; \(\tau\) denotes the relaxation time, \(\eta\) denotes the anticipation parameter, \(\kappa\) denotes the numerical stability parameter, \( \lambda_m\) denotes the number of lanes on link \(m\), and \(L_m\) denotes the length of link \(m\); \(V[\rho_{m,i}(k)]\) denotes the equilibrium speed when the link density is \(\rho_{m,i}(k)\) according to the fundamental diagram; \(q_o(k), d_o(k), w_o(k)\), and \(C_o\) denote the flow, demand, queue length, and capacity of the origin, respectively; \(q_s(k)\) denotes the flow in the off-ramp; finally, \(\rho_{m,i}(k)\), \(\rho_{\mathrm{crit},m}\) and \(\rho_{\max}\) denote the density, critical density, and jam density at link \(m\) segment \(i\) at time step \(k\).

\cref{eq:density_update} describes the link density update, while \cref{eq:speed_update} defines the link speed update. \cref{eq:flow} computes the flow as the product of density and speed.
\cref{eq:fundamental_diagram} is the fundamental diagram function. In METANET, origin links are modelled as store-and-forward links using a point-queue model. \cref{eq:queue_length_udpate} updates the queue length based on the conservation equation. \cref{eq:node_outflow} describes the outflow for on-ramps. For the outflow in the mainline source node, the second term in \cref{eq:node_outflow} must be removed. \cref{eq:queue_length_udpate} and \cref{eq:node_outflow} are unique to METANET and are not part of a second-order traffic model. 
For the on-ramp segment, an additional term, \cref{eq:on-ramp_speed}, is incorporated into \cref{eq:speed_update}.
\begin{equation} \label{eq:on-ramp_speed}
    - \frac{\delta T q_o(k) v_{m,1}(k)}{L_m \lambda_m (\rho_{m,1}(k) + \kappa)}
\end{equation}
where \(\delta\) denotes the merging parameter. When lane drop occurs on a segment, the resulting speed reduction is modelled as:
\begin{equation} \label{eq:lane_drop_speed}
    - \frac{\varphi T \Delta \lambda \rho_{m,N_m}(k)}{L_m \lambda_m \rho_{\textrm{crit},m}} v^2_{m,N_m}(k)
\end{equation}
where \(\varphi\) is a calibrated constant parameter. \(\Delta \lambda\) denotes the number of lanes dropped. For further details, readers may refer to \citet{messmerMETANET1990} and \citet{hegyiModel2004}.

\subsection{Bounded-METANET}

The core equation of METANET is \cref{eq:speed_update}, which governs speed in the next time step. The right-hand side of \cref{eq:speed_update} consists of three terms. The first term is the relaxation term, which describes the tendency of the speed to converge towards the equilibrium speed over the relaxation time~\(\tau\). The second term is the convection term, which describes the speed variations resulting from the incoming flow. A plausible explanation for this term is that drivers tend to maintain their speed relative to the flow entering a segment. The third term is the anticipation term, which accounts for speed decreases (or increases) when drivers anticipate an increase (or decrease) in density in the downstream segment. Notably, the anticipation term is a primary contributor to the capacity drop phenomenon within the METANET model.

\cref{eq:speed_update} does not inherently guarantee that the calculated speed remains within the physically reasonable range of zero to the maximum speed. To address the potential for unrealistic speed values, it is common practice to use a saturation function. 
\begin{equation}
    v_{m,i}^{\textrm{sat}}(k+1) = \begin{cases}
        v_{\max} & \textrm{if } v_{m,i}(k+1) > v_{\max} \\
        v_{m,i}(k+1) & \textrm{if } v_{\textrm{min}} < v_{m,i}(k+1) \leq v_{\max} \\
        v_{\textrm{min}} & \textrm{if } v_{m,i}(k+1) \leq v_{\textrm{min}}
    \end{cases}
\end{equation}
where \(v_{\textrm{min}}\) denotes a small positive constant speed assigned when the speed calculated by METANET becomes non-positive. Setting \(v_{\textrm{min}}\) to zero is typically avoided, because this would result in zero flow, leading to significant oscillations in both flow and speed that persist until stability is re-established. 

While the saturation function guarantees that the speed remains within the defined range, it does not inherently guarantee the reliability or robustness of the simulation outcomes. The primarily reason is the inherent unpredictability of when the saturation function will be invoked. For example, if the saturation function is triggered in scenarios matching the calibration data during the calibration process, subsequent simulation steps may align closely with real-world observations. However, when applying the calibrated METANET model to new scenarios, the activation of the saturation function introduces uncertainty: the resulting simulation trajectory may diverge from the expected behaviour. Such deviations occur frequently in practice, as the conditions triggering saturation depend on complex interactions among traffic state variables and calibrated parameters. Consequently, the downstream effects of saturation activation, such as long-term instability or unrealistic traffic dynamics, cannot be systematically evaluated or predicted.

The relaxation term in \cref{eq:speed_update}, a standard feature of second-order traffic models, is uniquely robust in this context. Unlike convection and anticipation terms, it does not produce unrealistic speed values due to its dependence on the fundamental diagram (\cref{eq:fundamental_diagram}), which inherently bounds speed through its equilibrium relationship with density. Using this property, we reformulate \cref{eq:speed_update} to develop a \textit{Bounded-METANET} model that inherently restricts speed to physically plausible ranges without relying on ad hoc saturation functions.

The reformulated link speed update equation in Bounded-METANET is as follows:
\begin{equation} \label{eq:refine_speed}
\begin{aligned}
    v_{m,i}(k+1) = & v_{m,i}(k) + \frac{T}{\tau} \left( V[\hat{\rho}_{m,i}(k)] - v_{m,i}(k) \right)
\end{aligned}
\end{equation}
where the virtual density \(\hat{\rho}_{m,i}(k)\) is calculated as:
\begin{equation} \label{eq:refine_density_hat}
   \hat{\rho}_{m,i}(k) = \rho_{m,i}(k) + \tilde{\eta} \frac{ \tilde{\kappa} }{\rho_{m,i}(k)+ \tilde{\kappa}} [\rho_{m,i+1}(k) - \rho_{m,i}(k)]
\end{equation}
where \( 0 \leq \tilde{\eta} \leq 1 \) is a weighting parameter, and \( \tilde{\kappa} > 0\) is a regularization constant introduced to avoid division by zero when \(\rho_{m,i}(k) = 0 \textrm{ veh/km} \).

The Bounded-METANET retains only the relaxation term from the original METANET speed update (\cref{eq:speed_update}), ensuring inherent speed stability. The anticipation effect, which is originally modelled through explicit downstream density gradients, is instead embedded within the fundamental diagram by substituting the actual density \(\rho_{m,i}(k)\) with the virtual density \(\hat{\rho}_{m,i}(k)\). This virtual density amplifies the perceived downstream density (via the term \(\tilde{\eta} \frac{ \tilde{\kappa} }{\rho_{m,i}(k)+ \tilde{\kappa}} [\rho_{m,i+1}(k) - \rho_{m,i}(k)]\) to mimic drivers’ anticipatory adjustments without risking unrealistic speed values.

The convection term, which links speed changes to upstream velocity gradients, is intentionally excluded. This omission aligns with empirical driver behaviour: speed adjustments are primarily responsive to downstream traffic conditions (e.g., congestion ahead) rather than upstream dynamics. By eliminating this term, the Bounded-METANET avoids introducing non-physical dependencies that could destabilize simulations.

When a segment includes on-ramps and lane drops, additional terms must be integrated into the Bounded-METANET model. Directly appending these terms to the virtual density calculation in \cref{eq:refine_density_hat}, however, risks generating unphysical values for \(\hat{\rho}_{m,i}(k)\). To address this, the merging and lane-drop effects are instead embedded into a modified virtual density \(\tilde{\rho}_{m,i}(k)\), ensuring stability while preserving the model’s bounded behaviour.

For segments with on-ramps, the merging effect is modelled as:
\begin{equation} \label{eq:refine_merging}
   \tilde{\rho}_{m,i}(k) = \hat{\rho}_{m,i}(k) + \tilde{\delta} \frac{ \tilde{\kappa} }{\rho_{m,i}(k)+ \tilde{\kappa}} \frac{ v_{m,i}(k) }{v_{\max}} \frac{ q_{o}(k) }{C_{o}} [\rho_{\textrm{max}} - \hat{\rho}_{m,i}(k)]
\end{equation}
where \(\tilde{\delta}\) (\( 0 \leq \tilde{\delta} \leq 1\)) denotes the merging parameter. $V[\hat{\rho}_{m,i}(k)]$ in \cref{eq:refine_speed} is replaced by $V[\tilde{\rho}_{m,i}(k)]$, scaling the perceived density by the on-ramp inflow 
\( q_{o}(k)\) relative to the ramp’s capacity \(C_{o}\). The term \(\frac{ \tilde{\kappa} }{\rho_{m,i}(k)+ \tilde{\kappa}} \) ensures numerical stability at low densities, while \(\rho_{\textrm{max}} - \hat{\rho}_{m,i}(k)\) bounds the adjustment to prevent oversaturation.

For segments with lane reductions, the density increase due to weaving phenomena is defined as
\begin{equation} \label{eq:refine_weaving}
   \tilde{\rho}_{m,i}(k) = \hat{\rho}_{m,i}(k) + \tilde{\varphi} \frac{ \Delta \lambda }{\lambda_m } \frac{ \rho_{m, N_m}(k) }{\rho_{\textrm{max}}} \frac{ v_{m,i}(k) }{v_{\max}}  [\rho_{\textrm{max}} - \hat{\rho}_{m,i}(k)]
\end{equation}
where \(\tilde{\varphi}\) (\( 0 \leq \tilde{\varphi} \leq 1 \)) is a constant parameter. \(\Delta \lambda\) is the number of lanes dropped, and \(\lambda_m \) is the total lanes in segment. Both density \(\rho_{m, N_m}(k)\) and speed \(v_{m,i}(k)\) are rescaled by the maximum density and maximum speed, respectively, ensuring the adjustment remains proportional to congestion severity.

When a segment features both an on-ramp and a lane drop, the combined virtual density modification is:
\begin{equation}~\label{eq:refine_merging_weaving}
   \tilde{\rho}_{m,i}(k) = \hat{\rho}_{m,i}(k) + \frac{1}{2} \left[ \tilde{\delta} \frac{ \tilde{\kappa} }{\rho_{m,i}(k)+ \tilde{\kappa}} \frac{ q_{o}(k) }{C_{o}} + \tilde{\varphi} \frac{ \Delta \lambda }{\lambda_m } \frac{ \rho_{m, N_m}(k) }{\rho_{\textrm{max}}} \right] \frac{ v_{m,i}(k) }{v_{\max}} [\rho_{\textrm{max}} - \hat{\rho}_{m,i}(k)]
\end{equation}

\cref{eq:refine_speed,eq:refine_density_hat,eq:refine_merging,eq:refine_weaving,eq:refine_merging_weaving} define the speed update mechanism in the Bounded-METANET model. Meanwhile, \cref{eq:density_update,eq:flow,eq:fundamental_diagram,eq:queue_length_udpate,eq:node_outflow} from the original METANET are retained in the Bounded-METANET formulation.

\subsection{Boundary analysis of the proposed model}

The Bounded-METANET model guarantees that link speeds and virtual densities remain within physically plausible bounds, as demonstrated below.

\subsubsection{Link speed boundaries}~\label{sec:bounds_speed}

Assume \( 0 \leq v_{m,i}(k) \leq v_{\max} \) at time step \(k\). By the definition of the fundamental diagram (\cref{eq:fundamental_diagram}), the equilibrium speed satisfies \( 0 \leq V[\hat{\rho}_{m,i}(k)] \leq v_{\max} \). Let \(\alpha := \frac{T}{\tau} \), where \(T\) is the simulation time step and \(\tau\) is the relaxation time. Since METANET typically assumes \(T \leq \tau\) to reflect finite driver adjustment rates, we have \( 0 \leq \alpha \leq 1 \). Substituting into \cref{eq:refine_speed}, the speed update becomes
\begin{equation}
    v_{m,i}(k+1) = v_{m,i}(k) + \alpha \left( V[\hat{\rho}_{m,i}(k)] - v_{m,i}(k) \right) = (1-\alpha) v_{m,i}(k) + \alpha V[\hat{\rho}_{m,i}(k)]
\end{equation}
Given \( 0 \leq \alpha \leq 1 \), the updated speed is a weighted average of the current speed and the equilibrium speed. Thus:
\begin{equation}
    0 \leq (1-\alpha) v_{m,i}(k) + \alpha V[\hat{\rho}_{m,i}(k)] \leq (1-\alpha) v_{\max} + \alpha v_{\max} = v_{\max} 
\end{equation}
Therefore, \( 0 \leq v_{m,i}(k+1) \leq v_{\max} \) holds for all \(k\), ensuring speed remains bounded without external saturation functions.

\subsubsection{Link density boundaries}

To ensure the reliability of the Bounded-METANET model, the virtual density \(\hat{\rho}_{m,i}(k)\) must remain within the physically plausible range \(0 \leq \hat{\rho}_{m,i}(k) \leq \rho_{\textrm{max}}\).

For the base virtual density in \cref{eq:refine_density_hat}, given \( \tilde{\kappa} > 0\) and \( 0 \leq \rho_{m,i}(k) \leq \rho_{\mathrm{max}} \), the term \(\frac{ \tilde{\kappa} }{\rho_{m,i}(k)+ \tilde{\kappa}} \) is bounded such that \(0 \leq \frac{ \tilde{\kappa} }{\rho_{m,i}(k)+ \tilde{\kappa}} \leq 1 \). Letting \(\beta := \tilde{\eta} \frac{ \tilde{\kappa} }{\rho_{m,i}(k)+ \tilde{\kappa}}\), where \(0 \leq \tilde{\eta} \leq 1\), it follows that \(0 \leq \beta \leq 1\). Substituting into \cref{eq:refine_density_hat}, the virtual density becomes:
\begin{equation}
    \hat{\rho}_{m,i}(k) = \rho_{m,i}(k) + \beta [\rho_{m,i+1}(k) - \rho_{m,i}(k)]
\end{equation}
This represents a convex combination of the current segment density \(\rho_{m,i}(k)\) and the downstream density \(\rho_{m,i+1}(k)\). Since both \(\rho_{m,i}(k)\) and \(\rho_{m,i+1}(k)\) lie within \([0,\rho_{\textrm{max}}] \), the weighted average \(\hat{\rho}_{m,i}(k)\) is also confined to \(0 \leq \hat{\rho}_{m,i}(k) \leq \rho_{\textrm{max}}\).

For the segments with on-ramps (\cref{eq:refine_merging}), the modified virtual density \(\tilde{\rho}_{m,i}(k)\) introduces an additive term scaled by parameters \(\tilde{\delta}\), \( \frac{ \tilde{\kappa} }{\rho_{m,i}(k)+ \tilde{\kappa}} \), \(\frac{ v_{m,i}(k) }{v_{\max}} \), \(\frac{ q_{o}(k) }{C_{o}} \), all of which are normalised to \( [0, 1]\). Consequently, the merging adjustment term satisfies:
\begin{equation}
    0 \leq \tilde{\delta} \frac{ \tilde{\kappa} }{\rho_{m,i}(k)+ \tilde{\kappa}} \frac{ v_{m,i}(k) }{v_{\max}} \frac{ q_{o}(k) }{C_{o}} [\rho_{\textrm{max}} - \hat{\rho}_{m,i}(k)] \leq \rho_{\textrm{max}} - \hat{\rho}_{m,i}(k)
\end{equation}
Thus, \(\tilde{\rho}_{m,i}(k)\) in \cref{eq:refine_merging} remains bounded within \([0, \rho_{\textrm{max}}]\).

A similar argument applies to lane-drop effects in \cref{eq:refine_weaving}. The parameters \(\tilde{\varphi}\), \( \frac{ \Delta \lambda }{\lambda_m } \), \( \frac{ \rho_{m, N_m}(k) }{\rho_{\textrm{max}}} \) and \(\frac{ v_{m,i}(k) }{v_{\max}} \) are each constrained to \([0, 1]\), ensuring the additive term does not exceed \(\rho_{\textrm{max}} - \hat{\rho}_{m,i}(k)\). The combined formulation in \cref{eq:refine_merging_weaving} further averages the merging and lane-drop contributions, preserving the bound.

\section{Numerical results}~\label{sec:numerical_results}

\subsection{CTM}

For comparative purposes, the first-order CTM \citep{daganzoCell1994} is incorporated as a baseline model. Our implementation follows the formulation by \citet{munozMethodological2004}. The flow conservation equation \cref{eq:density_update} is used in CTM as well. The flow in each cell is calculated by the minimum of the supplied flow from the upstream segment and receiving capacity based on a triangular fundamental diagram.
\begin{equation}
    q_{m,i}(k) = \min \left( S_{m,i}(k), R_{m,i+1}(k) \right)
\end{equation}
where \(S_{m,i}(k) = \min (v_{m,i}(k) \rho_{m,i}(k), C_{m,i})\) is the maximum flow that can be supplied by segment \(i\) link \(m\) in free-flow conditions, and \(R_{m,i}(k) = \min (C_{m,i}, u_i (\rho_{\max} - \rho_{m,i}(k)))\) is the maximum flow that can be received by downstream segment \(i\) link \(m\) in congested conditions. \(C_{m,i}\) denotes the capacity of segment \(i\) link \(m\) and \(u_i\) denotes the congestion wave speed. 

For merge segment involving on-ramps, 
\begin{equation}
    q_{m,i}(k) = \begin{cases}
S_{m,i}(k) & \text{if } S_{m,i}(k) + d_o(k) \leq R_{m,i+1}(k) \\
\max(0, R_{m,i+1}(k) - d_o(k)) & \text{if otherwise }
\end{cases}
\end{equation}
\begin{equation}
    q_{o}^{r}(k) = \begin{cases}
d_o(k) & \text{if } S_{m,i}(k) + d_o(k) \leq R_{m,i+1}(k) \\
R_{m,i+1}(k) - q_{m,i}(k) & \text{if otherwise } 
\end{cases}
\end{equation}
when \(S_{m,i}(k) + d_o(k) \geq R_{m,i+1}(k)\), the on-ramp flow \(q_o^r(k)\) is prioritized over the mainline flow \(q_{m,i}(k)\), ensuring ramp vehicles merge first into the constrained downstream capacity.

For diverge segments, the total exiting flow \(q_{m,i, \textrm{out}}(k)\) (mainline \(q_{m,i}(k)\) + off-ramp \(q_{\textrm{off}}(k)\)) is governed by
\begin{equation}
    q_{m,i, \textrm{out}}(k) = \min \left( S_{m,i}(k), \frac{R_{m,i+1}(k)}{1-\beta_i(k)} \right)
\end{equation}
where \(\beta_i(k)\) denotes the flow split ratio for the off-ramp. The off-ramp flow \(q_{\textrm{off}}(k) = \beta_i(k) q_{m,i, \textrm{out}}(k)\) and the mainline flow \( q_{m,i}(k) = (1 - \beta_i(k) ) q_{m,i, \textrm{out}}(k)\). \(\beta_i(k)\) denotes the flow split ratio for the off-ramp.

\subsection{Calibration method}~\label{sec:calibration}

Each model contains parameters that require calibration against reference data, which may originate from microscopic simulations or real-world field measurements. The calibration process involves minimizing discrepancies between model outputs (e.g., speed, density) and reference data using optimization techniques. To avoid manual weighting of variables, the Relative Root Mean Square Error (RRMSE) is adopted as the objective function instead of the standard RMSE. When reference data include segment-level speed and density, the objective function is defined as in \cref{eq:calibration}.

\begin{equation} \label{eq:calibration}
f = \frac{\mathrm{RMSE}_v}{\mathrm{Norm\_factor}_v } + \frac{\mathrm{RMSE}_{\rho}}{\mathrm{Norm\_factor}_{\rho}}
\end{equation}
where
\begin{align}
  \mathrm{RMSE}_v & = \sqrt{ \frac{1}{l K} \sum_{m=0}^{l} \frac{1}{N_m} \sum_{i=0}^{N_m} \sum_{k=0}^{K} \left(v_{m,i}^{\mathrm{model}}(k) - v_{m,i}^{\mathrm{ref}}(k) \right)^2 }, \label{eq:rmse_v} \\
  \mathrm{Norm\_factor}_v & = \sqrt{\sum_{m=0}^{l} \sum_{i=0}^{N_m} \sum_{k=0}^{K} \left(v_{m,i}^{\mathrm{ref}}(k) \right)^2 }. \label{eq:norm_v}
\end{align}
where \(\frac{\mathrm{RMSE}_v}{\mathrm{Norm\_factor}_v }\) and \( \frac{\mathrm{RMSE}_{\rho}}{\mathrm{Norm\_factor}_{\rho}} \) represent the Relative Root Mean Square Error (RRMSE) for speed and density, respectively. \(\mathrm{RMSE}_v\), and \(\mathrm{RMSE}_{\rho}\) denote the Root Mean Square Error (RMSE) for speed and density, respectively, while \( \mathrm{Norm\_factor}_v\), \(\mathrm{Norm\_factor}_{\rho} \) normalise these errors to unit less RRMSE values. \cref{eq:rmse_v} and \cref{eq:norm_v} show the calculation of \(\mathrm{RMSE}_v\) and \(\mathrm{Norm\_factor}_v\), respectively. Analogous expressions for density \(\mathrm{RMSE}_{\rho}\) and \(\mathrm{Norm\_factor}_{\rho}\) follow the same structure and are omitted for brevity. \(l\) denotes number of links in the network, \(N_m\) denotes the number of segments in link \(m\), \(K\) denotes the total simulation time steps. \(v_{m,i}^{\mathrm{model}}(k)\), and \(v_{m,i}^{\mathrm{ref}}(k) \) denote the speed at the time step \(k\) in segment \(i\) of link \(m\) from the model (e.g., METANET, Bounded-METANET, or CTM) and reference dataset, respectively. When the speed and flow of each segment are available in the reference dataset, the objective function is adjusted accordingly to incorporate the RRMSE for speed and flow.

The calibration is performed using the differential evolution algorithm \citep{stornDifferential1997}, a stochastic population-based optimisation method well-suited for global parameter tuning without requiring gradient information. The parameters to be calibrated and their typical limits are presented in \cref{tab:params_to_calibrate}. Minor adjustments are made in case studies to align with empirical observations or scenario-specific requirements, particularly for free-flow speed. A key benefit of the Bounded-METANET model is that all parameters (excluding those tied to the fundamental diagram, namely \(a_m, v_{\textrm{free}}, \rho_{\textrm{crit}}, \rho_{\textrm{max}}\)) can be straightforwardly bounded within intuitive ranges. Specifically, the bounds for \(\tilde{\eta}, \tilde{\kappa}\), and \(\tilde{\delta}\) remain consistent across scenarios, whereas the lower bound of \(\tau\) is determined by the simulation time step.

\begin{table}[tb]
\centering
\caption{Parameters to be calibrated in CTM, METANET and Bounded-METANET models}
\label{tab:params_to_calibrate}  %
\begin{tabular}{ c l c l }
\toprule
\textbf{Symbol} & \textbf{Description} & \textbf{Units} & \textbf{Limits} \\
\midrule
CTM & & & \\
$u$ & Congestion wave speed & km/h & [10, 30] \\
$v_{\textrm{free}}$ & Free-flow speed  & km/h & [90, 119] \\
$\rho_{\textrm{crit}}$  & Critical density & veh/km/lane  & [10, 30] \\
METANET & & & \\
$a_m$ & Fundamental diagram shape parameter  & - & [0.5, 2.5] \\
$v_{\textrm{free}}$ & Free-flow speed & km/h & [110, 200] \\
$\rho_{\textrm{crit}}$ & Critical density & veh/km/lane & [15, 40] \\
$\rho_{\max}$ & Maximum density & veh/km/lane & [110, 150] \\
$\tau$ & Relaxation time constant & s & [5, 80] \\
$\eta$ & Anticipation coefficient  & km$^2$/h & [10, 90] \\
$\kappa$ & Numerical stability parameter & veh/km/lane & [5, 50] \\
$\delta$ & Merging constant & - & [0.0001, 4] \\
Bounded-METANET & & & \\
$a_m$ & Fundamental diagram shape parameter & - & [0.5, 2.5] \\
$v_{\textrm{free}}$ & Free-flow speed & km/h & [110, 200] \\
$\rho_{\textrm{crit}}$ & Critical density & veh/km/lane & [15, 40] \\
$\rho_{\max}$ & Maximum density & veh/km/lane & [110, 150] \\
$\tau$ & Relaxation time constant & s & [5, 80] \\
$\tilde{\eta}$ & Anticipation weighting factor & - & [0, 1] \\
$\tilde{\kappa}$ & Numerical stability parameter & veh/km/lane & [1, 300] \\
$\tilde{\delta}$ & Merging weighting factor & - & [0, 1] \\
\bottomrule
\end{tabular}
\end{table}

\subsection{Case study 1: Calibration using the simulation dataset}

\begin{figure}[tb]
    \centering
    \medskip
    \begin{subfigure}[t]{.55\linewidth}
        \centering\includegraphics[width=.99\linewidth]{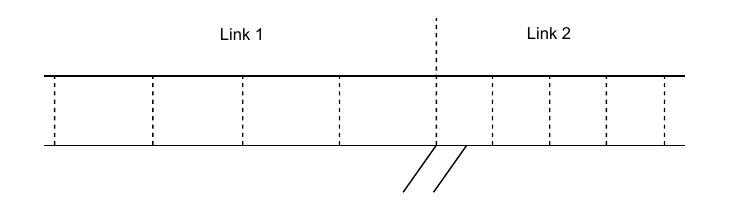}
        \caption{SUMO simulation network with two links and an on-ramp.} \label{fig:sumo_network}
    \end{subfigure}
    \begin{subfigure}[t]{.48\linewidth}
        \centering\includegraphics[width=.99\linewidth]{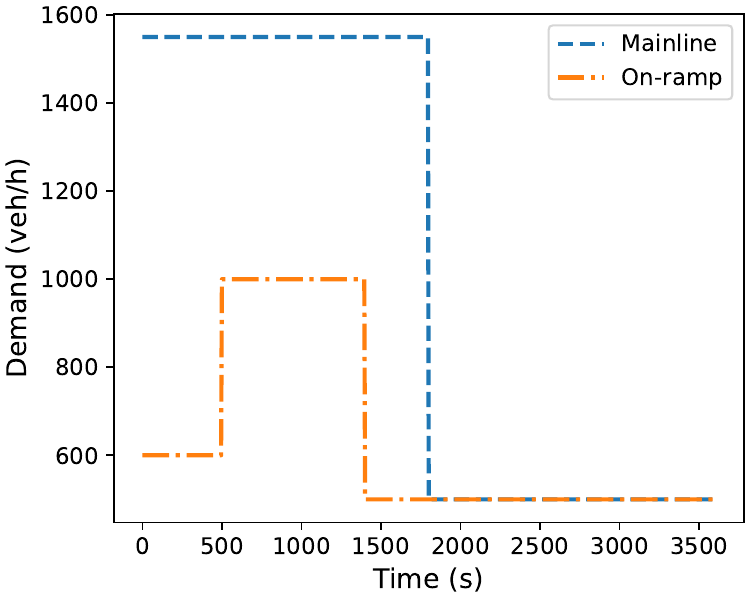}
        \caption{Demand profiles for mainline and on-ramp in the SUMO simulation.} \label{fig:sumo_demand}
    \end{subfigure}
    \caption{Network configuration and demand profiles in the SUMO simulation.} \label{fig:sumo_simulation}
\end{figure}

The traffic data generated by the SUMO simulation platform \citep{SUMO2018} are used for the calibration of the macroscopic model. The default Krauss car following model is used. To evaluate model performance, we compare three macroscopic flow models: the Cell Transmission Model (CTM), METANET, and the proposed Bounded-METANET. The simulated network, shown in \cref{fig:sumo_network}, comprises two main links and one on-ramp. Before merge, there are four segments, each with a length of 1 km. After merge, there are four segments, but each now has a length of 0.5 km. The demand profile in the SUMO simulation is shown in \cref{fig:sumo_demand}. A sharp peak in on-ramp demand exceeds the capacity of the merging area, leading to congestion.

The calibrated parameters derived using the methodology outlined in Section~\ref{sec:calibration} are summarized in \cref{tab:calibrated_params}. Model accuracy is quantified using the Root Mean Square Error (RMSE) for speed and density, as detailed in \cref{tab:rmse_sumo}. The results demonstrate that the second-order models, METANET and Bounded-METANET, significantly outperform the first-order CTM. This superiority stems from their ability to replicate capacity drops and dynamic speed transitions during congestion formation and dissipation. While CTM relies on steady-state equilibrium assumptions, it fails to represent the non-stationary traffic states inherent in this scenario. This limitation of CTM becomes pronounced when the on-ramp’s oversaturated demand triggers a sustained capacity drop. The proposed Bounded-METANET model achieves reductions in speed and density RMSE of over 40\% compared to the CTM. Moreover, it further improves upon the original METANET, reducing speed RMSE by 9.97\% and density RMSE by 17.11\%.

\begin{table}[tb]
\centering
\caption{Calibrated parameters in the case study 1 and 2}
\label{tab:calibrated_params}
\begin{tabular}{ c c r r }
\toprule
\textbf{Model} & \textbf{Symbol} & \makecell{\textbf{Calibrated values} \\ \textbf{in case study 1}} & \makecell{\textbf{Calibrated values} \\ \textbf{in case study 2}} \\
\midrule
\multirow{3}{*}{CTM} & $u$ & 16.15 & 16.83 \\
        & $v_{\textrm{free}}$ & 94.12 & 90.81 \\
        & $\rho_{\textrm{crit}}$ & 21.97 & 20.47 \\
\midrule
\multirow{8}{*}{METANET} & $a_m$ &  2.13 & 2.50 \\
    & $v_{\textrm{free}}$ &  118.99 & 130.60  \\
    & $\rho_{\textrm{crit}}$ &  29.41 & 22.85  \\
    & $\rho_{\max}$ &  143.78 & 138.76  \\
    & $\tau$ &  6.38 & 45.51 \\
    & $\eta$ &  23.29 & 87.75  \\
    & $\kappa$ &  6.64 & 7.95  \\
    & $\delta$ &  4.00 & 0.19  \\
\midrule
\multirow{8}{*}{Bounded-METANET} & $a_m$ & 1.81 & 1.92 \\
    & $v_{\textrm{free}}$ &  118.53 & 138.59 \\
    & $\rho_{\textrm{crit}}$ & 34.42 & 22.24 \\
    & $\rho_{\max}$ &  149.93 & 149.41 \\
    & $\tau$ &  12.35 & 80.00 \\
    & $\tilde{\eta}$ & 0.52 & 0.92  \\
    & $\tilde{\kappa}$ & 149.86 & 299.40  \\
    & $\tilde{\delta}$ & 0.45 & 1e-6 \\
\bottomrule
\end{tabular}
\end{table}

\begin{table}[tb]
\centering
\caption{Comparison of Root Mean Square Error (RMSE) for speed and density in three models in case study 1}~\label{tab:rmse_sumo}
\begin{tabular}{c c c c c}
\toprule
\textbf{Model} & \textbf{RMSE for Speed (km/h)} & \textbf{RMSE for Density (veh/km)} \\
\midrule
CTM & 23.38 & 13.01  \\
METANET & 15.35 & 9.35 \\
Bounded-METANET & \textbf{13.82} & \textbf{7.75} \\
Reduction vs CTM & 40.89\% & 40.43\% \\
Reduction vs METANET & 9.97\% & 17.11\% \\
\bottomrule
\end{tabular}
\end{table}

\begin{figure}[tb]
    \centering
    \medskip
    \begin{subfigure}[t]{.48\linewidth}
        \centering\includegraphics[width=.99\linewidth]{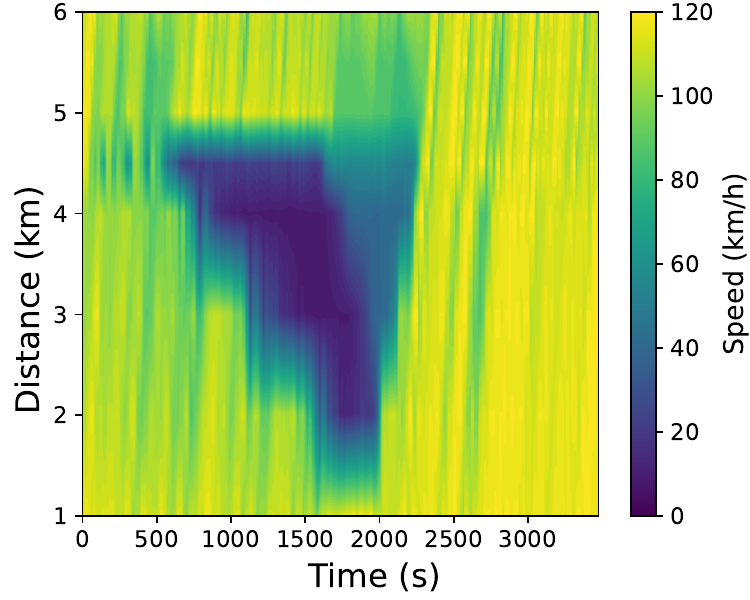}
        \caption{Speed across segments in SUMO simulation (reference data).} \label{fig:sumo_speed}
    \end{subfigure}
    \begin{subfigure}[t]{.48\linewidth}
        \centering\includegraphics[width=.99\linewidth]{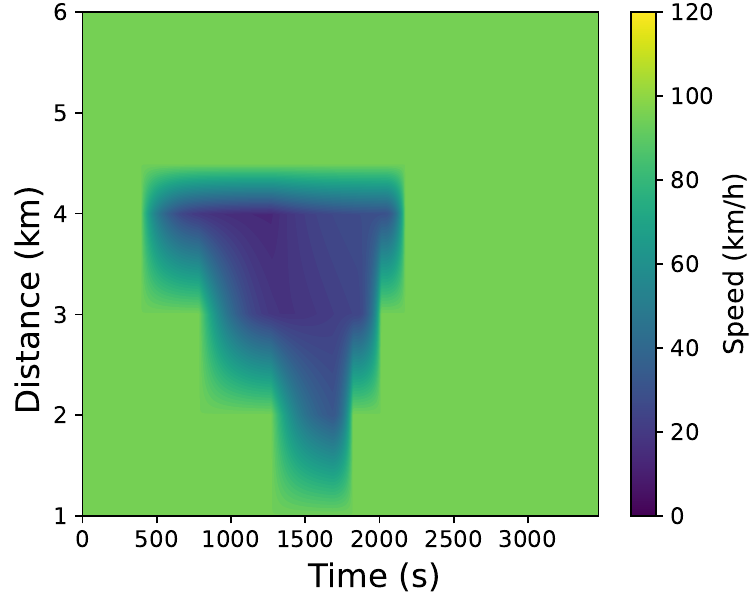}
        \caption{Speed across segments in CTM simulation.} \label{fig:ctm_speed}
    \end{subfigure}
    
    \begin{subfigure}[t]{.48\linewidth}
        \centering\includegraphics[width=.99\linewidth]{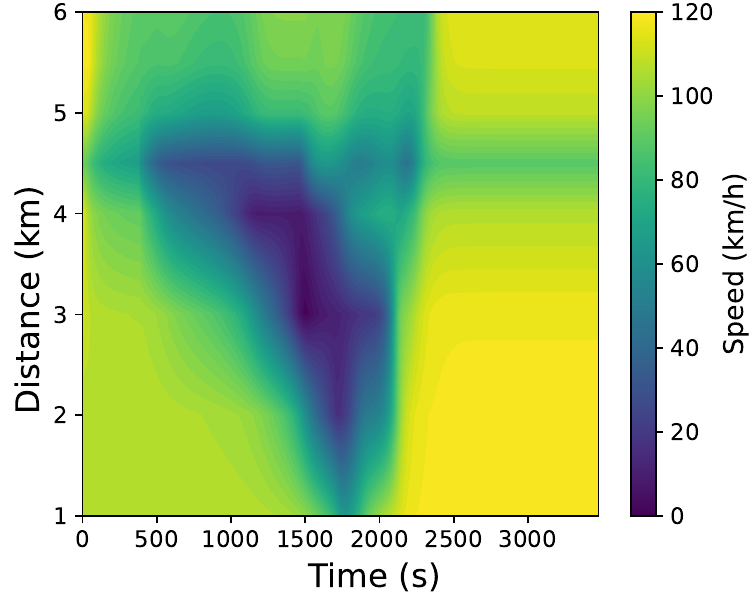}
        \caption{Speed across segments in METANET simulation.} \label{fig:metanet_speed}
    \end{subfigure}
    \begin{subfigure}[t]{.48\linewidth}
        \centering\includegraphics[width=.99\linewidth]{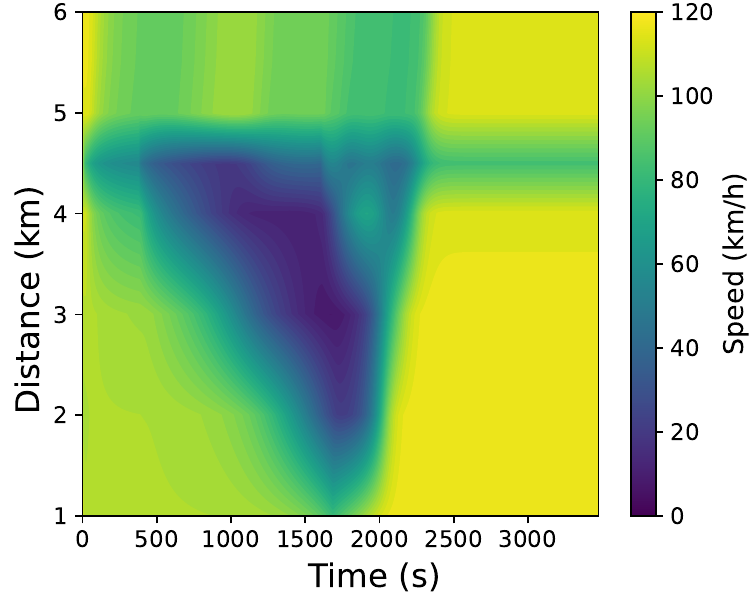}
        \caption{Speed across segments in Bounded-METANET simulation.} \label{fig:refine-metanet_speed}
    \end{subfigure}
    \caption{Contour plot of speed (km/h) across segments and time in case study 1, comparing SUMO reference data with CTM, METANET, and Bounded-METANET simulations. }~\label{fig:speed_results}
\end{figure}

The speed profiles in each segment from the three macroscopic models and reference data generated by SUMO simulations are shown in \cref{fig:speed_results}. All models successfully reproduce the traffic shockwave triggered by high demand from the on-ramp and its upstream propagation. However, significant differences emerge in their representation of free-flow dynamics and congestion evolution. The CTM enforces a constant free-flow speed in uncongested conditions and fails to capture non-equilibrium speed adjustments outside the congested area, such as approaching and leaving the congested area. This rigidity forces the calibrated free speed to assume unrealistically low values to approximate observed traffic behaviour. Furthermore, the lack of capacity drop in the CTM leads to an abrupt abrupt congestion dissipation in the merging area (at 4 km), which is inconsistent with the congested merging segment and the gradual recovery observed in SUMO. Both second-order models replicate congestion build-up at the merge and subsequent speed drops, aligning more closely with SUMO than CTM. However, METANET exhibits sharper speed transitions in congested states, resulting in spatially confined low-speed regions. In contrast, Bounded-METANET produces smoother speed transitions and broader congestion zones, better matching the gradual speed decrease and extended recovery phases seen in SUMO. This behaviour demonstrates that Bounded-METANET has better performance in modelling the severe congestions.

\begin{figure}[tbp]
    \centering
    \medskip
    \begin{subfigure}[t]{.48\linewidth}
        \centering\includegraphics[width=.99\linewidth]{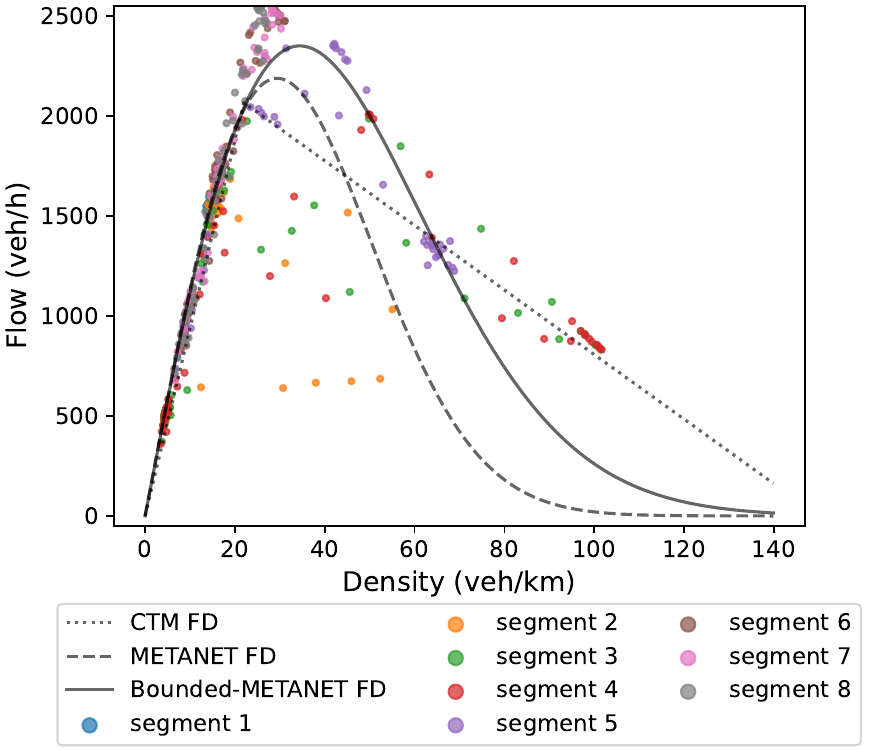}
        \caption{Flow-density relationship in SUMO simulation (reference data) with overlaid fundamental diagrams of CTM, METANET, and Bounded-METANET.} \label{fig:sumo_fd}
    \end{subfigure} 
        \begin{subfigure}[t]{.48\linewidth}
        \centering\includegraphics[width=.99\linewidth]{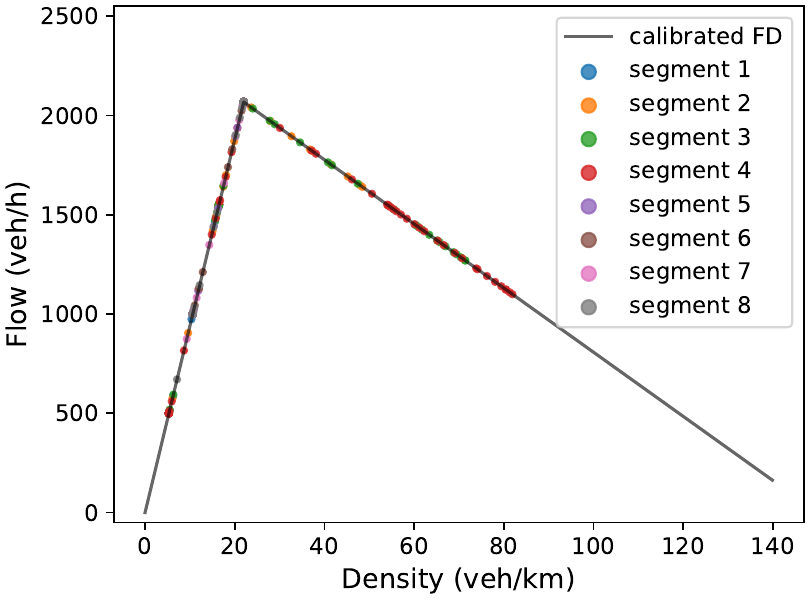}
        \caption{Flow-density relationship in CTM simulation.} \label{fig:ctm_fd}
    \end{subfigure} 
    \begin{subfigure}[t]{.48\linewidth}
        \centering\includegraphics[width=.99\linewidth]{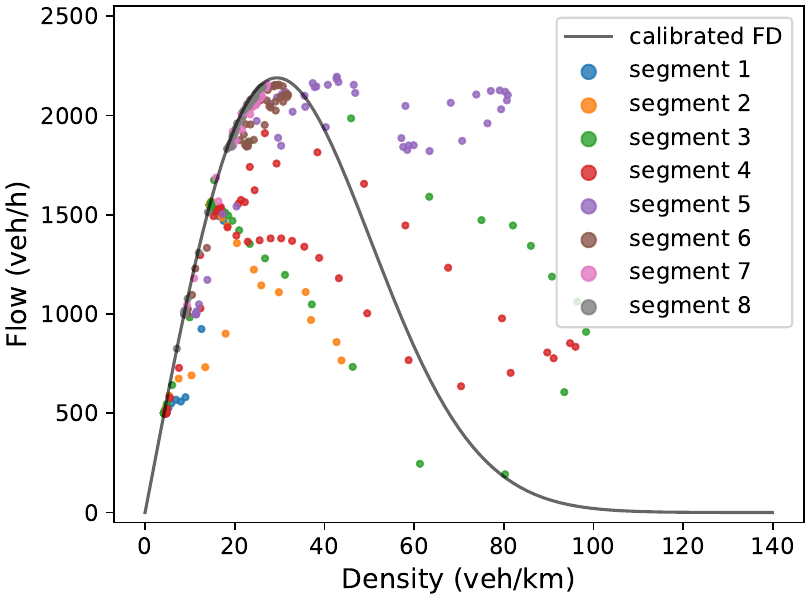}
        \caption{Flow-density relationship in METANET simulation.} \label{fig:metanet_fd}
    \end{subfigure}
    \begin{subfigure}[t]{.48\linewidth}
        \centering\includegraphics[width=.99\linewidth]{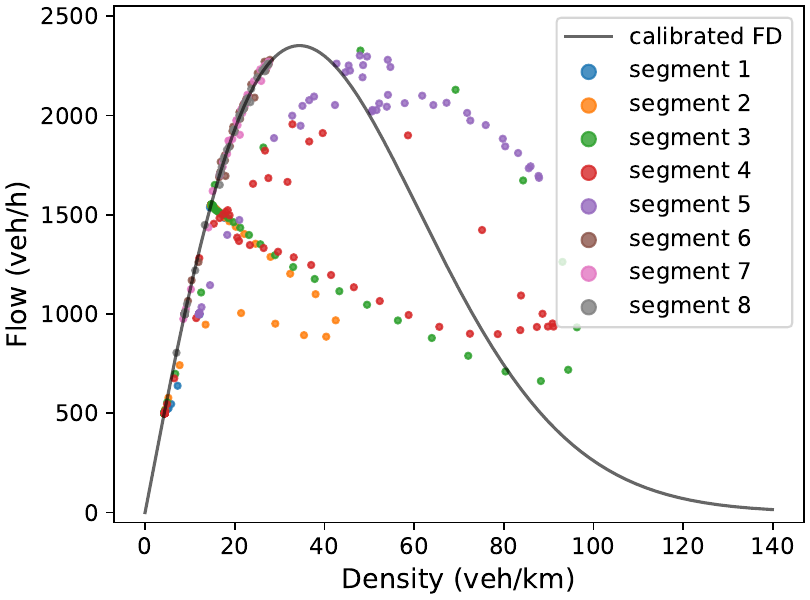}
        \caption{Flow-density relationship in Bounded-METANET simulation.} \label{fig:bounded_fd}
    \end{subfigure}
    \caption{Flow-density scatter plots and calibrated fundamental diagrams for Case Study 1, comparing SUMO reference data with CTM, METANET, and Bounded-METANET simulations.}~\label{fig:result_fd}
\end{figure}

To gain further insight into model performance, \cref{fig:result_fd} compares the calibrated fundamental diagrams of CTM, METANET, and Bounded-METANET against traffic states derived from SUMO simulations (data aggregated at 1-minute intervals). The Bounded-METANET fundamental diagram (\cref{fig:sumo_fd}) aligns more closely with the reference data than METANET, which exhibits reduced capacity and lower congested-phase flows. This discrepancy arises because METANET cannot fully capture speed reductions, due to limitations in its anticipation and merging terms. To compensate, the model artificially lowers congested-phase capacity during calibration. Such adjustments reflect METANET’s high sensitivity to fundamental diagram parameters under congestion. In contrast, Bounded-METANET maintains consistency with the reference fundamental diagram, demonstrating its capacity to simulate congestion propagation without distorting the equilibrium flow-density relationship. CTM, meanwhile, exhibits the lowest critical density and capacity among the models, along with a relatively high jam density (\cref{fig:result_fd}). This suggests that CTM artificially amplifies the congestion representation by compressing the free-flow regime and inflating the congested regime, which is a consequence of its steady-state equilibrium assumption. \cref{fig:metanet_fd} and \cref{fig:bounded_fd} highlight key differences in non-equilibrium traffic state modelling. METANET generates flow-density points that deviate from the fundamental diagram even in free-flow conditions, which is a result of its convection and anticipation terms perturbing the relaxation dynamics. Conversely, Bounded-METANET’s flow-density points adhere closely to the fundamental diagram in free-flow regimes, as its speed update relies solely on the relaxation term. This adherence underscores the bounded model’s stability and its avoidance of unphysical perturbations inherent in METANET’s additional terms.

\begin{figure}[tb]
  \centering
  \medskip
  \begin{subfigure}[t]{.45\linewidth}
      \centering\includegraphics[width=.99\linewidth]{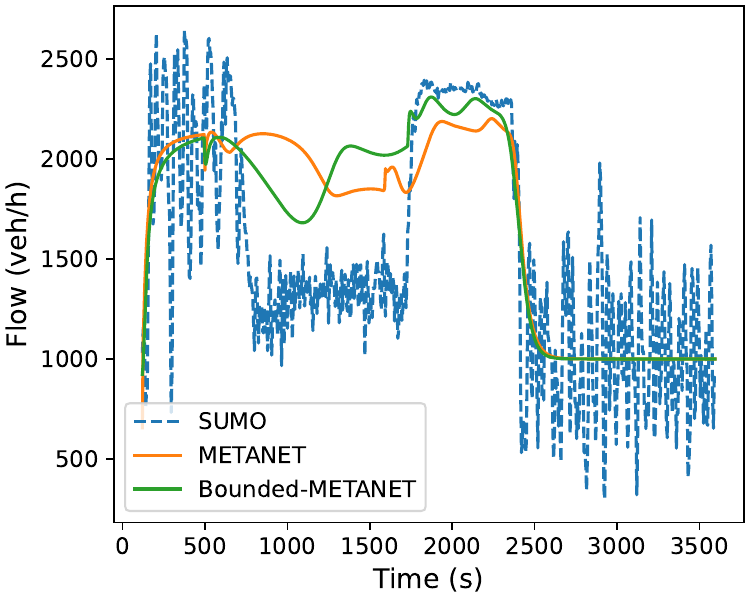}
      \caption{Flow after merge} \label{fig:after_merge_flow_sumo}
  \end{subfigure}
  \begin{subfigure}[t]{.45\linewidth}
      \centering\includegraphics[width=.99\linewidth]{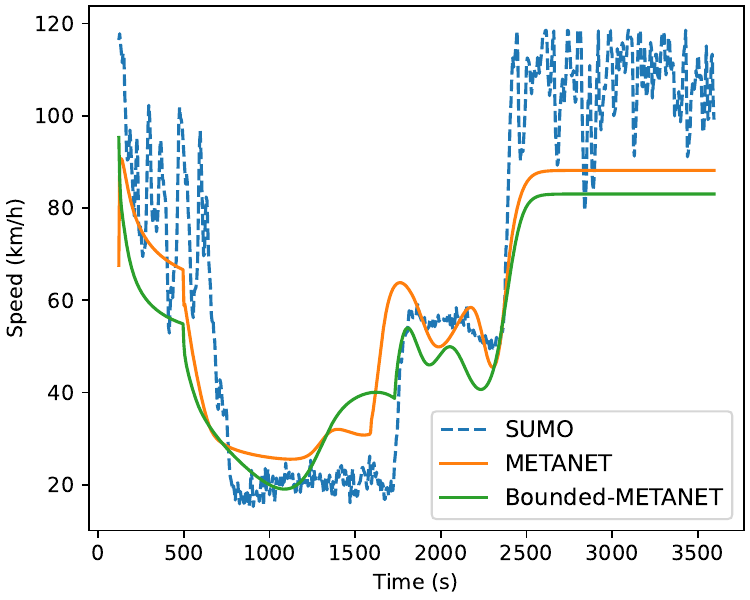}
      \caption{Speed after merge.} \label{fig:after_merge_speed_sumo}
  \end{subfigure}
  \caption{Flow and speed in the merging segment in the SUMO (used as reference data), METANET and Bounded-METANET.}~\label{fig:after_merge_sumo}
\end{figure}

To investigate the models’ ability to simulate capacity drop, the flow and speed dynamics in the first segment of link 2 after the merging are shown in \cref{fig:after_merge_sumo}. Both METANET and Bounded-METANET capture capacity drop phenomena to comparable degrees, though neither aligns precisely with the flow reduction in the reference data. This discrepancy likely arises from the strong capacity drop in the SUMO simulation, where flow drops sharply from \numrange{2000}{1300}~\si{veh\per\hour}, representing a reduction exceeding 35\%. Notably, in Bounded-METANET, the onset of flow drop occurs earlier than in METANET, and recovery to maximum capacity is faster. These differences can be partially attributed to the convection term in METANET. Specifically, the initial appearance of congestion in the merging segment triggers a speed reduction due to increased density. However, the speed in the final segment of link 1 remains relatively high, and the positive convection term in METANET delays the speed decrease. Conversely, during recovery from congestion to maximum capacity, low upstream speeds further delay METANET's speed recovery. In the Bounded-METANET, the flow exhibits two distinct stages. Initially, it exhibits a significant reduction, followed by a gradual increase to an intermediate level. Even during this second stage, the flow remains below the maximum capacity, indicating persistent capacity drop. Notably, the maximum flow achieved after recovery in Bounded-METANET exceeds that of METANET, showing a pattern consistent with the calibrated fundamental diagram. Speed profiles in \cref{fig:after_merge_speed_sumo} are similar across both models. After 2500 s, speeds return to initial low-demand levels but remain significantly below free-flow speed despite matching reference flow values (118~km/h in reference data). This suggests that traffic density after merging persists at higher levels than observed empirically.

\begin{figure}[tb]
    \centering
    \includegraphics[width=.8\linewidth]{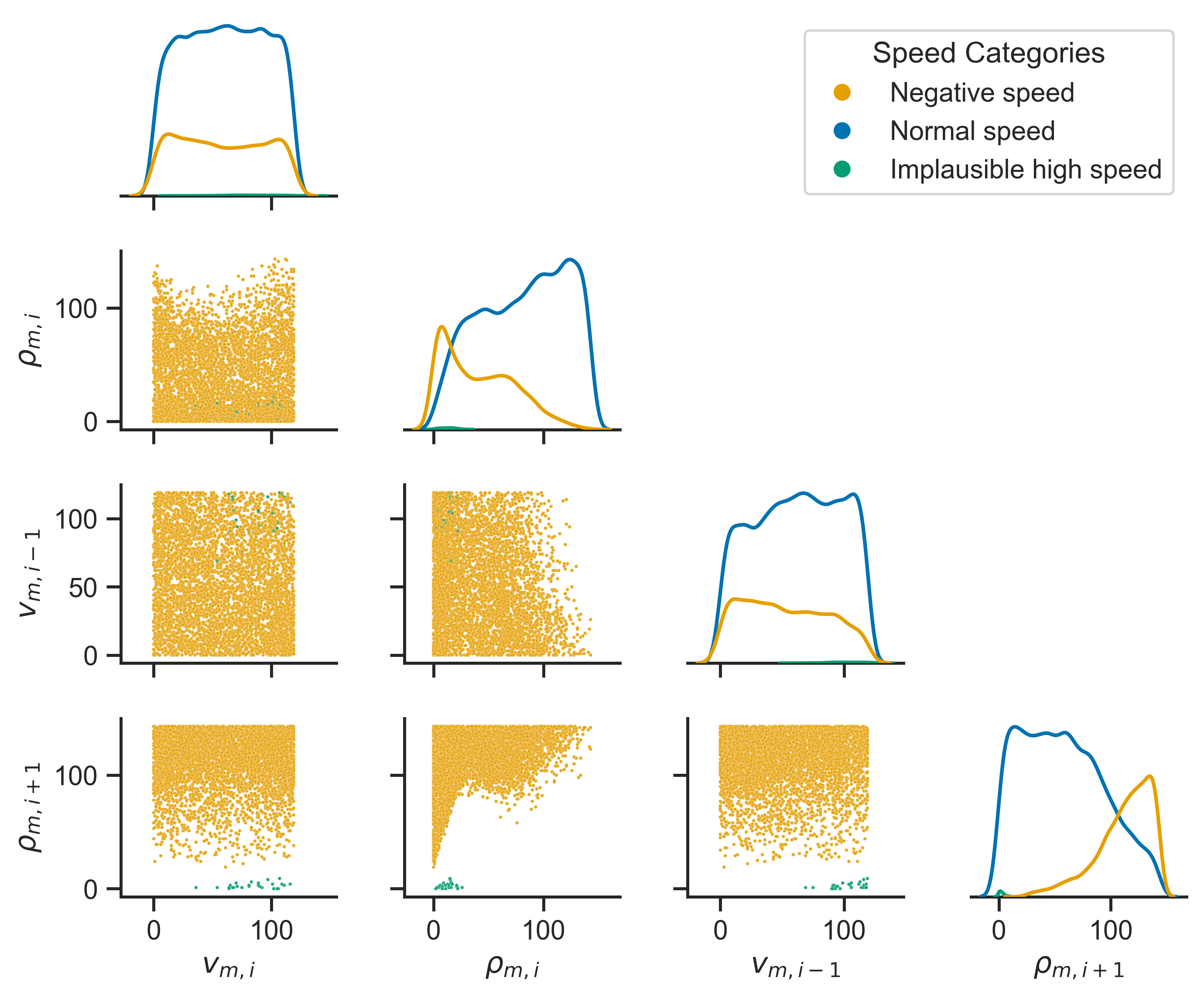}
    \caption{Speed distribution of the calibrated METANET model using the SUMO dataset. Diagonal subplots show the marginal distribution of three speed categories: negative speed, normal speed and implausible high speed. Off-diagonal subplots show only the categories of negative speed and implausible high speed to highlight calibration anomalies.} \label{fig:speed_range}
\end{figure}

To critically examine the limitations of the METANET model in capturing speed dynamics, we analyse speed distributions derived from calibrated parameters. During calibration, strong penalties are imposed to eliminate non-physical solutions - specifically, negative speeds and speeds exceeding free-flow conditions. This ensures that speeds remain constrained within physically plausible bounds in the calibration process. The speed update mechanism (\cref{eq:speed_update}) reveals that the subsequent time step speed depends on four state variables: current segment speed \(v_{m,i}\), upstream segment speed \(v_{m,i-1}\), current segment density \(\rho_{m,i}\), and downstream segment density \(\rho_{m,i+1}\). To quantify the model’s behaviour, we evaluate all possible combinations of these variables across their operational ranges, discretised at 1 km/h increments for speed and 1 veh/km increments for density. This exhaustive combinatorial analysis generates a complete distribution of speeds for the subsequent time step. A representative subset of 20,000 data points, constituting merely 0.006\% of the total parameter space, was randomly sampled to illustrate the distribution in \cref{fig:speed_range}. Despite this sparse sampling, the results exhibit consistent distributional properties. Speeds are categorized into three classes: negative values (\(<\) 0 km/h), normal speeds (0 km/h to free-flow speed), and implausible high speed (\(>\)free-flow speed). Analysis reveals that 73.66\% of simulated speeds fall within the normal range, while 26.20\% yield non-physical negative values. Notably, only 0.14\% exceed the free-flow speed threshold, highlighting the model's inherent preference for under-prediction over over-prediction, particularly in unstable or congested regimes.

The results confirm empirical observations: while careful calibration ensures plausible speed bounds during the calibration process, METANET offers no inherent guarantees for retaining these bounds during real-world deployment. Specifically, the model exhibits a non-negligible probability of generating unrealistic speeds - particularly negative values, which occur in approximately 25\% of simulated time steps. While implausibly high speeds are less frequent, they remain a critical concern, especially in optimisation contexts where such anomalies could destabilise control policies. \cref{fig:speed_range} reveals distinct dependencies for these anomalies:
\begin{itemize}
    \item Negative speeds:
        \begin{itemize}
            \item \textbf{Weak dependence} on current speed (\(v_{m,i}(k)\)) and upstream speed (\(v_{m,i-1}(k)\)).
            \item \textbf{Strong dependence} on low local density ( \(\rho_{m,i}(k)\) ) and high downstream density ( \(\rho_{m,i+1}(k)\) ), reflecting overcompensation in the anticipation term under steep density gradients.
        \end{itemize}
    \item Implausibly high speeds: 
        \begin{itemize}
            \item \textbf{Primary driver:} low downstream density (\(\rho_{m,i+1}(k)\)).
            \item \textbf{Secondary drivers:} low current density (\(\rho_{m,i}(k)\)), high upstream speed (\(v_{m,i-1}(k)\)) and high current speed (\(v_{m,i}(k)\)).
        \end{itemize}
\end{itemize}
These dependencies highlight METANET’s sensitivity to abrupt density transitions, which can destabilise its speed update mechanism despite calibration.

\subsection{Case study 2: Calibration using the real dataset}

\begin{figure}[tb]
    \centering
    \medskip
    \includegraphics[width=.8\linewidth]{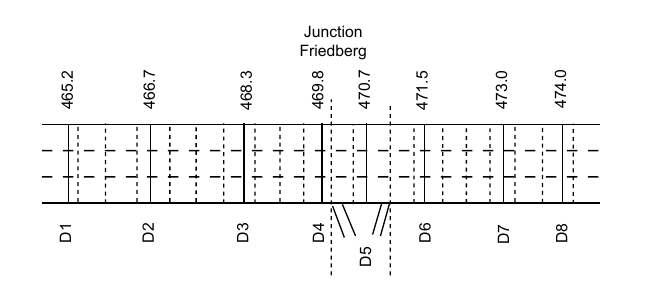}
    \caption{Network layout and segmentation scheme for the German motorway A5-South study area. The roadway is partitioned into three homogeneous links between an off-ramp and on-ramp pair. Solid vertical lines indicate loop detector locations (D1-D8), while dashed lines indicate segment boundaries.} \label{fig:german_road}
\end{figure}

\begin{figure}[tbh]
    \centering
    \medskip
    \begin{subfigure}[t]{.48\linewidth}
        \centering\includegraphics[width=.99\linewidth]{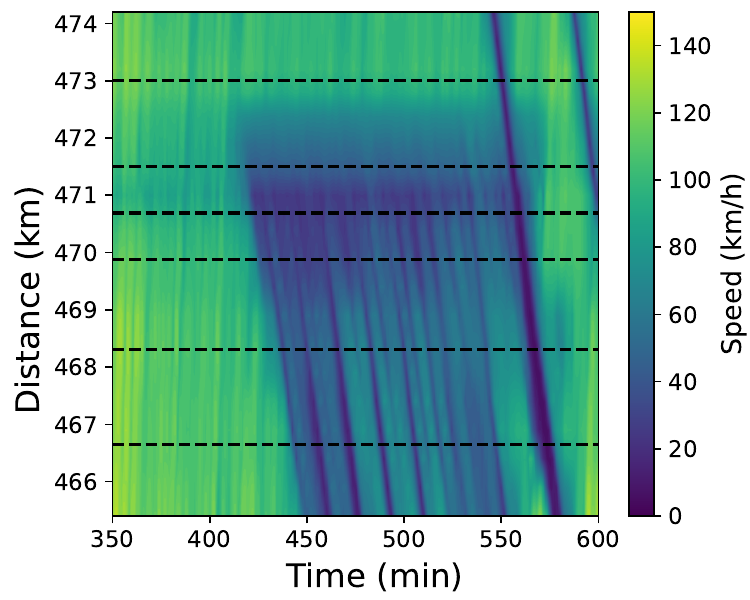}
        \caption{Speed in the German A5-South dataset.} \label{fig:german_speed}
    \end{subfigure}
        \begin{subfigure}[t]{.48\linewidth}
        \centering\includegraphics[width=.99\linewidth]{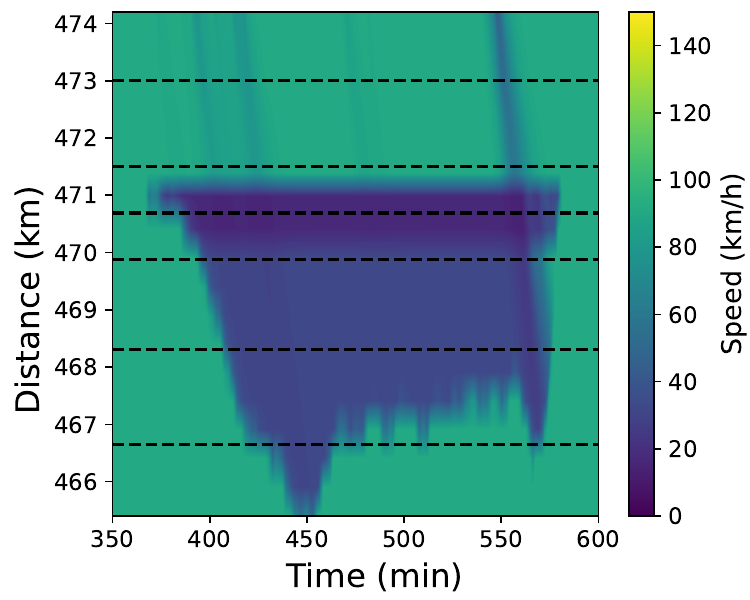}
        \caption{Speed in the CTM simulation.} \label{fig:german_ctm_speed}
    \end{subfigure}
    \begin{subfigure}[t]{.48\linewidth}
        \centering\includegraphics[width=.99\linewidth]{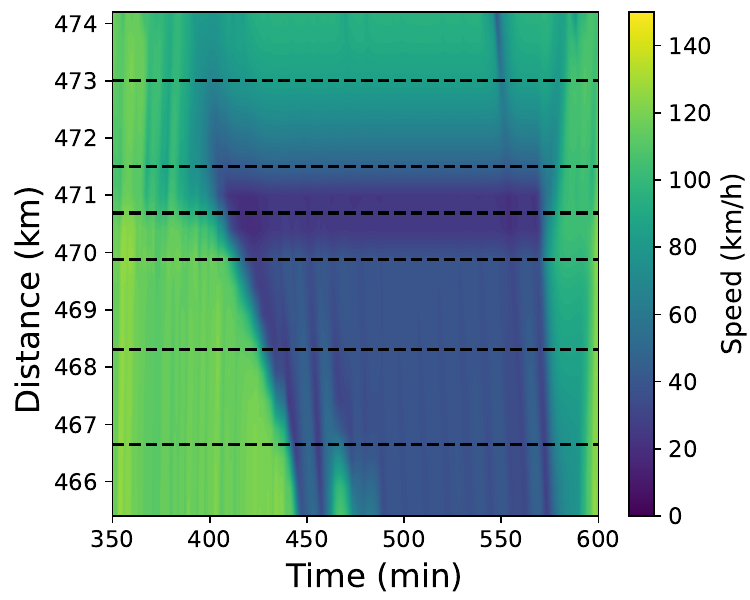}
        \caption{Speed in the METANET simulation.} \label{fig:german_metanet_speed}
    \end{subfigure}
    \begin{subfigure}[t]{.48\linewidth}
        \centering\includegraphics[width=.99\linewidth]{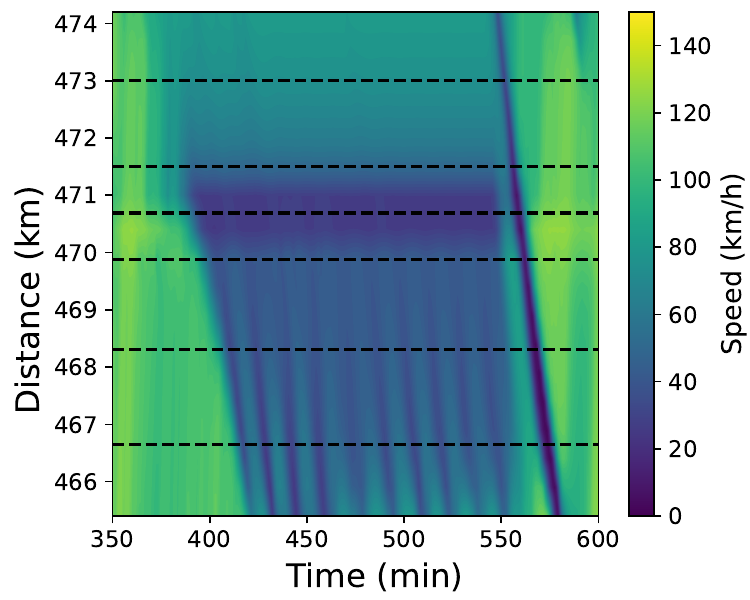}
        \caption{Speed in the Bounded-METANET simulation.} \label{fig:german_refine-metanet_speed}
    \end{subfigure}
    \caption{Speed profiles in each segment for case study 2: Observed data compared to CTM, METANET, and Bounded-METANET simulations.}~\label{fig:speed_results_german}
\end{figure}

The data collected from loop detectors located on a section of the German motorway A5-South near Frankfurt on 28th May 2001 are used in this study. 
As illustrated in \cref{fig:german_road}, the study network features an off-ramp and on-ramp pair, which divides the roadway into three links to preserve homogeneity in road geometry within each segment. Speed and flow measurements were recorded at 1-minute intervals, distinguishing between passenger cars and trucks. Trucks constituted 15-20\% of the observed traffic composition. The raw dataset exhibited substantial noise, which required preprocessing through the Adaptive Smoothing Method (ASM) \citep{treiberReconstructing2002} using parameters specified in \cref{tab:asm_params}. \cref{fig:german_speed} presents the smoothed speed data, which serves as the reference dataset for model calibration. Analysis indicates that congestion formation on the mainline primarily originates from increased demand at the on-ramp situated between detectors D5 and D6 during peak periods.

\begin{table}[tb]
\centering
\caption{Parameters in the adaptive smoothing method (ASM)}
\label{tab:asm_params}
\begin{tabular}{ c l c }
\toprule
\textbf{Symbol} & \textbf{Description} & \textbf{Values} \\
\midrule
$\sigma$ & Range of spatial smoothing & 0.5 km \\
$\tau$   & Range of temporal smoothing & \SI{0.5}{min} \\
$c_{\textrm{free}}$ & Propagation speed of perturbations in free traffic & \SI{70}{km/h}  \\
$c_{\textrm{cong}}$ & Propagation speed of perturbations in congested traffic & \SI{-15}{km/h} \\
$V_c$   & Crossover from free to congested traffic & \SI{50}{km/h} \\
$\Delta V$ & Width of the transition region  & \SI{10}{km/h} \\
\bottomrule
\end{tabular}
\end{table}

As density data were not directly measured, the calibration objective function uses Relative Root Mean Square Errors (RRMSE) for speed and flow. Following the methodology outlined in \cref{sec:calibration}, parameter estimates for the three models are summarised in \cref{tab:calibrated_params}, with corresponding RMSE values for speed and flow provided in \cref{tab:rmse_german}. Second-order models exhibit superior speed modelling performance than first-order models, attributed to their incorporation of momentum conservation principles to capture nonequilibrium traffic dynamics. METANET achieves the lowest speed RMSE (18.64 km/h), while Bounded-METANET yields the smallest flow error (710.23 veh/h). Notably, the CTM demonstrates unexpectedly competitive flow error (801.11 veh/h) despite its simpler first-order structure. A performance trade-off emerges between the second-order models: Bounded-METANET increases speed RMSE by 16.36\% compared to METANET but reduces flow RMSE by 14.65\%, suggesting METANET over-prioritizes speed accuracy at the expense of flow accuracy. Bounded-METANET addresses this imbalance, achieving corresponding reduction in flow error (14.65\% vs. METANET) while maintaining reasonable speed modelling performance (21.69 km/h RMSE). These results highlight the ability of the proposed Bounded-METANET model to better balance speed-flow interdependencies under identical calibration constraints.

\begin{table}[tb]
\centering
\caption{Comparison of Root Mean Square Error (RMSE) for speed and flow across three models in case study 2. The RMSE for flow is calculated using the total flow in three lanes.}
\label{tab:rmse_german}
\begin{tabular}{c c c c c}
\toprule
\textbf{Model} & \textbf{RMSE for speed (km/h)} & \textbf{RMSE for flow (veh/h)} \\
\midrule
CTM & 27.81 & 801.11  \\
METANET & \textbf{18.64} & 832.17 \\
Bounded-METANET & 21.69 & \textbf{710.23} \\
Reduction vs CTM & 22.01\% & 11.34\% \\
Reduction vs METANET & -16.36\% & 14.65\% \\
\bottomrule
\end{tabular}
\end{table}

The calibrated speed contours from the three models against empirical data from the German A5-South are shown in \cref{fig:speed_results_german}. There are two shockwaves starting at the downstream of the study network and propagating upstream due to the accident on the motorway. All models are capable of replicating the shockwave starting from the merging area because of the high demand at the on-ramp. Consistent with Case Study 1, CTM underestimates free-flow speeds in uncongested regions. Additionally, it fails to fully propagate congestion to the upstream boundary, resulting in a truncated congested zone compared to empirical observations. While METANET improves congestion modelling accuracy, its shockwave propagation speed exceeds empirical observations by a significant margin. This discrepancy likely contributes to METANET’s high flow RMSE (832.17 veh/h; \cref{tab:rmse_german}), and illustrates a critical limitation: optimising speed RMSE alone does not guarantee a physically plausible traffic state reconstruction. Bounded-METANET achieves superior spatio-temporal accuracy, closely matching the empirical shockwave propagation speed. It is the only model that accurately reproduces the widespread shockwave propagating from the downstream to upstream boundary within the study area. It can be concluded that under noisy observational conditions, METANET may produce inaccurate results regarding shockwave speed, whereas Bounded-METANET better balances these competing objectives by maintaining speed-flow accuracy and preserving physically realistic shockwave behaviour.

\begin{figure}[tbp]
    \centering
    \medskip
    \begin{subfigure}[t]{.48\linewidth}
        \centering\includegraphics[width=.99\linewidth]{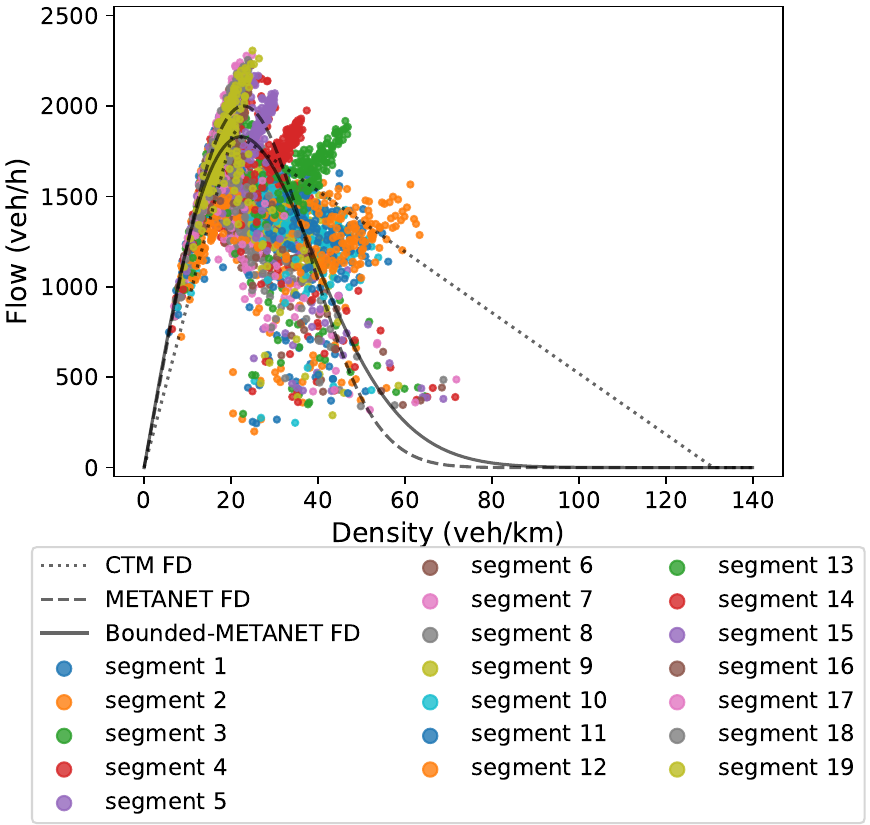}
        \caption{German A5-South dataset and three calibrated fundamental diagrams in three macroscopic models.} \label{fig:german_fd}
    \end{subfigure}
        \begin{subfigure}[t]{.48\linewidth}
        \centering\includegraphics[width=.99\linewidth]{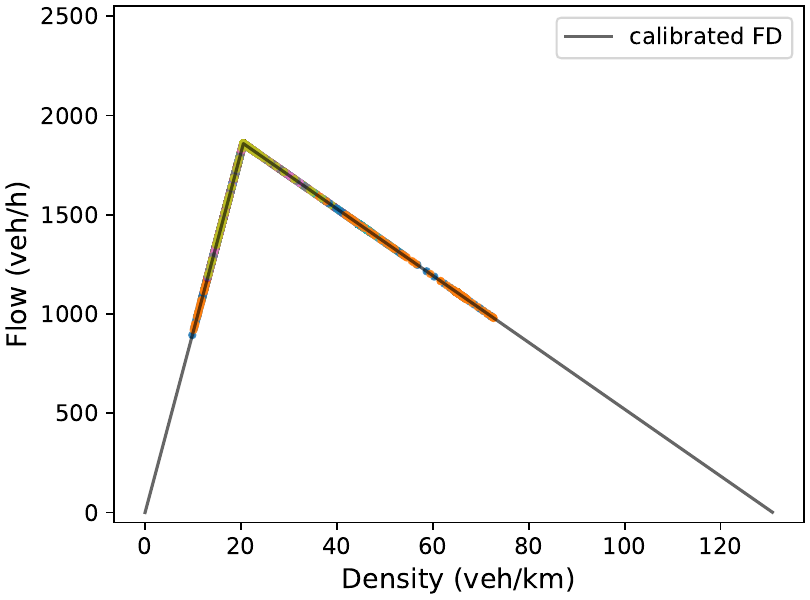}
        \caption{CTM: Simulated flow-density data with calibrated FD.} \label{fig:german_fd_ctm}
    \end{subfigure} \\
    \begin{subfigure}[t]{.48\linewidth}
        \centering\includegraphics[width=.99\linewidth]{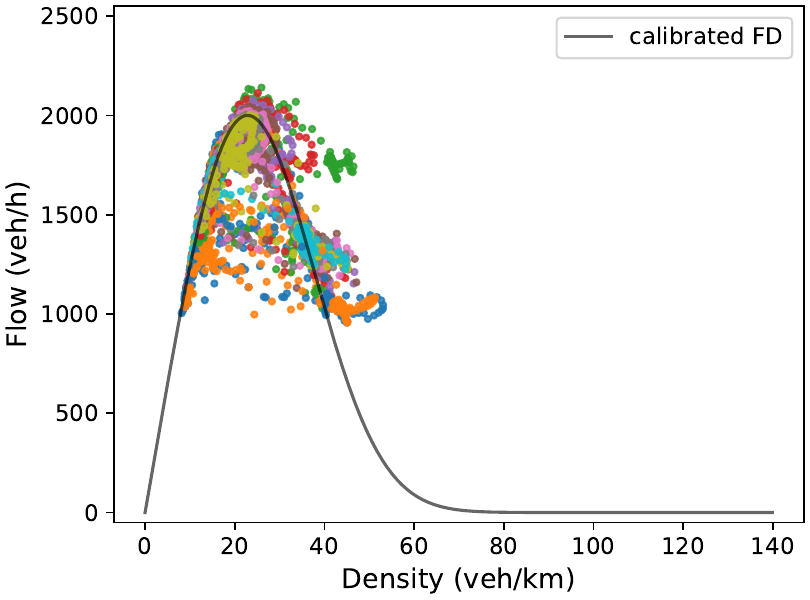}
        \caption{METANET: Simulated flow-density data with calibrated FD.} \label{fig:german_fd_metanet}
    \end{subfigure}
    \begin{subfigure}[t]{.48\linewidth}
        \centering\includegraphics[width=.99\linewidth]{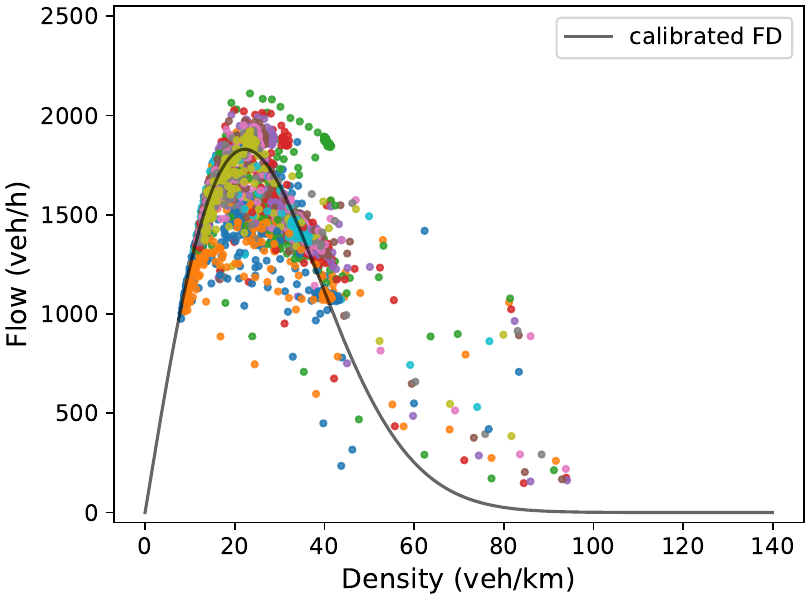}
        \caption{Bounded-METANET: Simulated flow-density data with calibrated FD.} \label{fig:german_fd_bounded}
    \end{subfigure}
    \caption{Flow-density relationships in Case Study 2. (a) Empirical data; simulated results from (b) CTM, (c) METANET, and (d) Bounded-METANET. Calibrated fundamental diagrams are overlaid on scatter plots of simulated data points.}~\label{fig:result_fd_german}
\end{figure}

The empirical and model-calibrated fundamental diagrams (FDs) for the three traffic models are compared in \cref{fig:german_fd}. Despite applying the Adaptive Smoothing Method (ASM), empirical noise persists in the data. All three calibrated FDs exhibit nearly identical critical densities. While the capacities of CTM and Bounded-METANET closely align, METANET predicts a higher capacity but underestimates flows at densities exceeding 40 veh/km compared to Bounded-METANET. The CTM’s triangular FD deviates significantly from the empirical data in congested states, which is a likely contributor to its elevated speed and flow errors. METANET also struggles to reproduce flows below 1000 veh/h or densities above 60 veh/km, regions where Bounded-METANET better approximates the empirical distribution. These FD discrepancies likely explain Bounded-METANET’s superior flow RMSE during calibration.

\begin{figure}[tb]
  \centering
  \medskip
  \begin{subfigure}[t]{.45\linewidth}
      \centering\includegraphics[width=.99\linewidth]{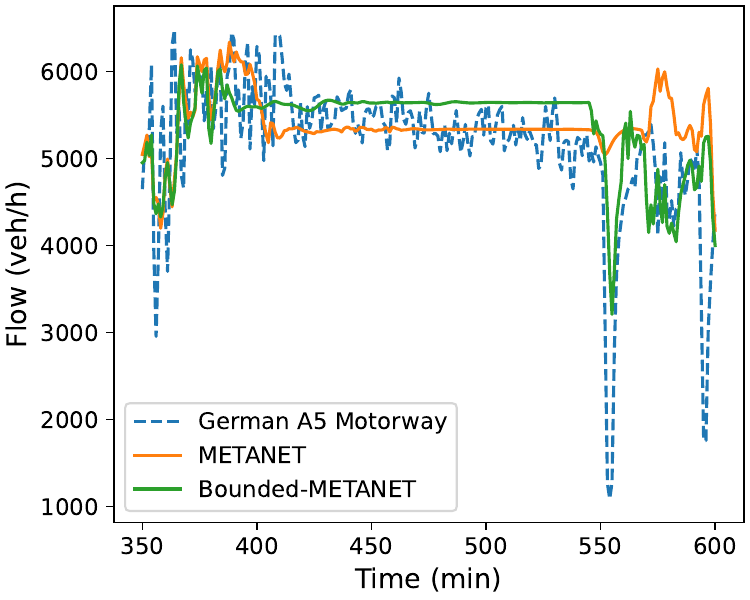}
      \caption{Flow after merge} \label{fig:after_merge_flow_germany}
  \end{subfigure}
  \begin{subfigure}[t]{.45\linewidth}
      \centering\includegraphics[width=.99\linewidth]{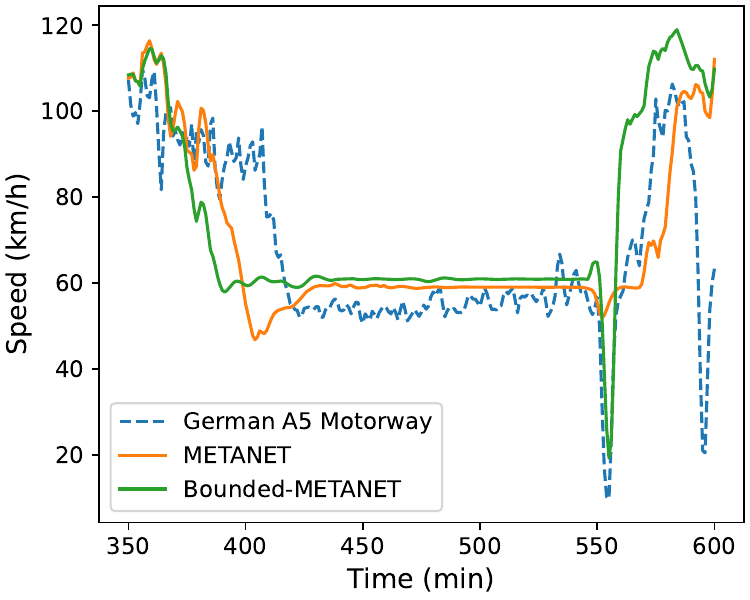}
      \caption{Speed after merge.} \label{fig:after_merge_speed_germany}
  \end{subfigure}
  \caption{Flow and speed after merge in the study of German A5-South.}~\label{fig:after_merge_germany}
\end{figure}

\cref{fig:after_merge_germany} compares simulated and observed flow-speed dynamics downstream of a merging segment for METANET and Bounded-METANET. The empirical data exhibit two notable features: (1) a significant drop in both flow and speed after 420 minutes, (2) two abrupt drops in flow and speed near the 550 minutes and 600 minutes, attributed to congestion propagating upstream from downstream bottlenecks. METANET closely replicates the magnitude and timing of the initial flow reduction (at 420 minutes) after merging, which aligns well with empirical trends. In contrast, Bounded-METANET underestimates the severity of the initial flow drop, maintaining marginally higher flow rates than observed after the disturbance. Both models show early speed reductions compared to empirical data. A key divergence occurs in the last 50 minutes (550-600 minutes): Bounded-METANET better captures the late-stage flow and speed collapse, particularly the first sharp decrease, whereas METANET fails to reproduce this behaviour. In conclusion, METANET outperforms Bounded-METANET in replicating the flow drop after merge because of the capacity drop (at 420 minutes), although this is not reflected in the overall simulation performance.

\section{Conclusions}~\label{sec:conclusions}

This study introduces Bounded-METANET, a discrete-time second-order macroscopic traffic flow model designed to resolve critical limitations of the classical METANET model. By reformulating METANET’s speed dynamics, the proposed model rigorously enforces physically meaningful speed bounds, eliminating non-physical predictions (e.g., negative speeds or speeds exceeding free speed) that arise in the original formulation. This new formulation enhances both computational safety (e.g., avoiding divergent states) and practical usability in traffic management systems. The bounds for the parameters are also easier to determine for calibration. Validated against SUMO-generated data and empirical loop detector measurements, Bounded-METANET demonstrates robust performance alongside benchmark models (METANET and CTM), with distinct advantages in critical traffic regimes.

The conclusions are summarised as follows.

1) Both Bounded-METANET and METANET outperform the first-order CTM in modelling traffic flow because of their ability of modelling non-equilibrium traffic states. While CTM excels in simplicity, stability, and congestion propagation accuracy, its reliance on equilibrium FD assumptions limits fidelity in complex scenarios.

2) In noise-free (but still stochastic) SUMO simulations, Bounded-METANET achieves the lowest RMSE for speed and density. With real-world noisy data, METANET yields superior speed RMSE, whereas Bounded-METANET better approximates flow. However, METANET produces unrealistic shockwave speeds, leading to flow errors that exceed even those of CTM. This highlights the need for METANET calibration methods that balance the trade-offs between speed and flow error.

3) METANET more accurately captures the flow drop after merging due to capacity drops than Bounded-METANET, but its FD is distorted severely under congestion. Bounded-METANET achieves more accurate FD in high-congestion regimes, reliably simulating both non-equilibrium states and the congestion propagation, as evidenced by two case studies.

To strengthen the robustness and applicability of Bounded-METANET, evaluation under diverse operational scenarios is critical. This includes testing under adverse weather conditions, recurrent congestion (e.g., lane-drop bottlenecks), and non-stationary demand patterns. Additionally, validation on large-scale motorway networks using empirical trajectory or loop detector data would confirm its scalability and generalisability. While Bounded-METANET demonstrates preliminary capability in modelling capacity drops at merging sections, persistent discrepancies between its predictions and reference data (in both simulated and empirical datasets) highlight opportunities for refinement. Future work should investigate whether refinement of merging modelling can improve accuracy in these scenarios. Finally, the practical impact of Bounded-METANET should be quantified through integration into traffic management frameworks. For example, embedding the model into model predictive control (MPC) systems for ramp metering or variable speed limits could demonstrate its advantages in improving stability, avoiding non-physical states, and enhancing control robustness compared to classical METANET or CTM.

\section*{Acknowledgements}

This research is  partly funded by the Australian Research Council (ARC) through Dr. Mehmet Yildirimoglu's Discovery Early Career Researcher Award (DECRA; DE220101320). We sincerely thank Dr. Martin Treiber (TU Dresden) for generously providing the German Highway A5 dataset in this study.

\bibliographystyle{elsarticle-harv}
\bibliography{references}

\end{document}